\documentclass[pss]{wiley2sp} 
\usepackage{amsmath}
\usepackage{amssymb}

\usepackage{epsfig,psfrag}
\usepackage{dcolumn}
\usepackage{graphicx}
\usepackage[pdf]{pstricks}
\usepackage{verbatim}
\usepackage{widetext}
\usepackage[off]{auto-pst-pdf}

\begin{document}

\title{Iterative path integral summation for nonequilibrium quantum transport}
\titlerunning{Iterative summation of path integrals}

\author{%
  S. Weiss\textsuperscript{\Ast \textsf{\bfseries 1}},
  R. H\"utzen\textsuperscript{\textsf{\bfseries 2}},
  D. Becker\textsuperscript{\textsf{\bfseries 3}}, 
  J. Eckel\textsuperscript{\textsf{\bfseries 2}},
  R. Egger\textsuperscript{\textsf{\bfseries 2}}, 
  M. Thorwart\textsuperscript{\textsf{\bfseries 4}}}

\authorrunning{S. Weiss et al.}

\mail{e-mail
  \textsf{weiss@thp.uni-due.de}, Phone:+49-203-379-2969}

\institute{%
  \textsuperscript{1}\,Theoretische Physik, Universit\"at Duisburg-Essen and 
CENIDE, Lotharstr.1, 47048 Duisburg, Germany\\
  \textsuperscript{2}\,Institut f\"ur Theoretische Physik IV,
Heinrich-Heine-Universit\"at D\"usseldorf, Universit\"atsstr.1, 40225
D\"usseldorf, Germany\\
   \textsuperscript{3}\,Departement Physik, Universit\"at Basel, 
Klingelbergstrasse 82, 4056 Basel, Switzerland\\
     \textsuperscript{4}\,I. Institut f\"ur Theoretische Physik, Universit\"at
Hamburg, Jungiusstra{\ss}e 9, 20355 Hamburg, Germany\\}

\received{XXXX, revised XXXX, accepted XXXX} 
\published{XXXX} 


\abstract{
\abstcol{
We have developed a numerically exact approach to 
compute real-time path integral expressions for 
quantum transport problems out of equilibrium.  
The scheme is based on a deterministic iterative summation of 
the path integral (ISPI) for the generating
function of nonequilibrium observables of interest, e.g., the charge current or
dynamical quantities of the central part. 
Self-energies due to the leads, being nonlocal in time, 
are fully taken into account within a finite memory time, 
thereby including non-Markovian effects. Numerical results are
extra\-polated first to vanishing (Trotter) time discretization
and, second, to infinite memory time.  The method is applied to 
nonequilibrium transport through a single-impurity Anderson dot in the first
place. We benchmark our results in various 
regimes of the rich parameter space. 
In the respective regime of validity, ISPI results are shown to match those of other state-of-the art methods.
Especially, we have chosen the mixed valence regime of the Anderson model to compare ISPI to time 
dependent density matrix renormalization group (tDMRG) and functional RG calculations.
Secondly, we determine the nonequilibrium current $I(V)$ 
through a molecular junction in presence of a vibrational mode. We have found an
exact mapping of the single impurity Anderson-Holstein}{ model to an effective
spin-1 problem. In analytically tractable regimes, 
as the adiabatic phonon or weak molecule-lead coupling
regime, we reproduce known perturbative results. 
Studying the crossover regime between those limits shows that the  Franck-Condon
blockade persists in the quantum limit. At low temperature,
the Franck-Condon steps in the  $I(V)$ characteristics are smeared due to 
nonequilibrium conditions. The third system under investigation here is the
magnetic Anderson model which consists of a spinful single-orbital quantum
dot with an incorporated quantum mechanical spin-1/2 magnetic impurity. Coulomb
interaction together with the exchange coupling of the magnetic impurity with the electron spins
strongly influence the dynamics. We investigate the  
nonequilibrium tunneling current through the system as a function of exchange and Coulomb interaction as well as 
the real-time  impurity polarization. From the real-time evolution of physical 
observables, we are able to determine characteristics
of the time-dependent nonequilibrium current and the relaxation dynamics of the
impurity.  These examples illustrate that the ISPI technique is particularly
well suited for the deep quantum regime, 
when all time and energy scales are of the same order of magnitude.}
}

\maketitle   
\section{Introduction} 
Small condensed matter systems that behave as transistors have attracted numerous research activities in recent years. 
When placed in a transport setup, current and noise measurements are used to determine the properties of the
usually strongly correlated quantum mechanical system.  
In as well as out of thermal equilibrium, several quantum many-body properties
of systems such as quantum dots or single molecule junctions  
are accessible in experiments \cite{Heinzel,molel}. 
By down-scaling electronic transistors to the nanometer scale, quantum dots (QD's)
allow to control electronic and/or spin 
properties of single electrons and might be used to impress and read out 
 information. Single molecules, inheriting also  
vibrational degrees of freedom, are ideal candidates for such devices
since they can rather easily be functionalized such that current switches are
 implemented in a nanoscale environment 
\cite{molel1,Nitzan07,Weber02,Ruitenbeek97,Scheer04,Park}. 
Furthermore, a controlled design of functional groups attached to molecules seems in reach. The  field of 
molecular electronics is at the intersection of interdisciplinary research areas, combining chemical and physical properties of molecules.
Under nonequilibrium conditions transport through molecules challenges experimentalists as well as theoreticians. 
Nonequilibrium in this 
review is referred to as the situation that the Fermi energies, provided by
source and drain electrodes and located 'left' and 'right' of the structure, are
not equal such that electrons are transferred through the system by tunneling. 
The voltage is assumed to drop at the nanostructure.  
In such finite voltage bias situations, many interesting physical effects
arise due to the quantum nature of the electrons and their strong Coulomb
interaction among each other and/or by their interaction with molecular vibrations or local magnetic moments. Prominent examples are resonant
tunneling or the Kondo effect \cite{Koenig,Meir,Hershfield}. Different techniques are used to approach the 
strong coupling limit. For instance,  the regime of  low energy, low temperature/bias has been approached by Fermi 
liquid theory \cite{Oguri01},  via interpolative schemes \cite{Aligia},  using integrability concepts \cite{Konik02}, 
or by the perturbative renormalization group (RG) \cite{Kaminski00,Rosch04}. 
A perturbative RG analysis has been performed 
by Schoeller and K\"onig \cite{Schoeller}.  The nonequilibrium generalization of Wilson's numerical 
RG  approach  \cite{Costi1} and of the fRG method \cite{Jakobs} have been discussed. 
Transport features have also been discussed by perturbation theory in the interaction strength \cite{Fujii}. On the other hand, perturbatively 
treating the tunneling matrix elements is a powerful way of describing the incoherent regime, see Ref.~\cite{Koenig}. 
Density matrix renormalization group techniques have been extended to the nonequilibrium regime \cite{Schmitteckert,Schollwoeck}. 
In addition the flow equation method allows to study the Toulouse point of the Kondo model \cite{Lobaskin}. 

In case that  a molecule is placed between electrodes,  an appropriate theoretical description is needed to deal with its characteristic vibrational 
(phonon) degrees of freedom \cite{molel,pederson}. The interplay between mechanical and 
electronic degrees of freedom is of interest in other areas of physics as well, e.g., for inelastic tunneling spectroscopy \cite{nitzan}, 
nanoelectromechanical systems \cite{nems},  break junctions \cite{jan}, and suspended semiconductor or carbon-based nanostructures  \cite{Steele09,adrian2,ensslin,HuettelWitkamp}.  Including vibrational degrees of freedom via a simple Anderson Holstein (AH)
model, where a spinless electronic level is coupled to a  single oscillator
mode, shows several effects as the 
Franck-Condon blockade, negative differential conductance, or current induced
heating or cooling \cite{Mitra,WangThoss2009}; 
for a review, see Ref.~\cite{nitzan}. As for the interacting Anderson model, 
analytical approaches typically address different corners of 
parameter space, a full theory that connects those corners seems 
not in reach at present. 

A third topic adressed in this review is the investigation of a magnetic QD. Those setups have been studied experimentally in ensembles which are
particularly suited for the investigation by laser and electromagnetic fields
\cite{mackowski:3337,PhysRevB.71.035338,0295-5075-81-3-37005,PhysRevLett.102.177403,Zutic2009,PhysRevB.81.245315,Ochsenbein2009}. They are
designed with standard lithographic methods and are technologically well
established. Moreover, embedding individual magnetic Mn ions into quantum dots
and studying the electrical properties is possible \cite{PhysRevLett.97.107401,PhysRevLett.98.106805,Hanson2008}. Small quantum dots with few charge carriers
and a single magnetic impurity may become important candidates for efficient
high density spintronic devices. 

There is a considerable need for numerical methods which describe small quantum
systems out of equilibrium accurately and, ideally, treat simple model
Hamiltonians numerically exactly. Numerical renormalization
group \cite{cornaglia} or quantum Monte Carlo (QMC) calculations 
\cite{arrachea,RMP,Oguri95,Wang,lothar,schiro} provide a possible line of attack to those problems.  
Due to the dynamical sign problem, these calculations become increasingly difficult at low temperatures, but in several 
parameter regions, the stationary steady-state regime seems accessible. 
Based on non-standard ensembles, the steady state is described in a recent work by Han \cite{Han}, using an 
imaginary-time QMC approach followed by a double analytical continuation scheme. This last step is numerically the most difficult part \cite{dirks}.

We here review the development of a novel numerical scheme denoted as 
{\sl iterative summation of real-time path integrals}\ (ISPI), in order to 
address quantum transport problems out of equilibrium\cite{ispi}.
Many-body systems driven out of equilibrium are known
\cite{Kaminski00,Abanin05,Mitra05b} to acquire a steady state that may
be quite different in character from their ground-state properties.
Details of the steady state may depend on the nature of the
correlations, as well as on the way in which the system is driven out
of equilibrium. 
Our ISPI approach, described in detail below, provides an alternative and numerically exact method to 
tackle out-of-equilibrium transport in correlated quantum dots. 
Based on the evaluation of the full 
nonequilibrium Keldysh generating function, along with the inclusion of  suitable source
terms, observables of interest are computed. 
 It builds on the fact that nonlocal in time correlations, induced by the fermionic leads,  
 decay exponentially in the long-time limit at any 
finite temperature. Within a characteristic time $\tau_{c}$,  all correlations are taken into account,  while 
for larger times, the correlations 
are dropped due to their exponentially small contributions. This allows us to construct an iterative scheme to
evaluate the generating  function. An appropriate 
extrapolation procedure allows to eliminate the 
Trotter time discretization error 
(the Hubbard-Stratonovich (HS) transformation 
below requires to discretize time) as well as  
the finite memory-time error. This  yields  the desired numerically exact value for
the observables of interest. Note that the need for a finite memory time makes our approach difficult to apply at very low energies ($T,V\to 0$).  Fortunately, other
methods are available in this regime. At finite $T$ or $V$, the requirement of not too long memory times is exploited and the spin path summation remains tractable. 
Recently, also Segal et al., see Ref.~\cite{segal} have provided an alternative formulation of the ISPI approach in terms of Feynman Vernon like influence functionals. 

The ISPI scheme is implemented here for three characteristic impurity models. First, the single-level Anderson impurity model 
\cite{Anderson,Flensberg,Tsvelik,Schiller,Horvath}, second the spinless Anderson Holstein model \cite{nitzan}, which mimics the behavior of a 
molecular quantum dot.  Finally, when magnetic molecules are investigated, additional couplings to localized spin impurities are 
important. Results for the relaxation dynamics of such a system are presented in this article as well.

The present paper is organized as follows. In Sec.\ \ref{sec:Model}, we 
introduce the model for the quantum dot coupled to normal 
leads. We present the computation of the generating  
function for the nonequilibrium Anderson model there as well. 
The presence of an external source term allows to 
calculate the current as a functional derivative. In Sec.\ \ref{sec:ISPI} 
we introduce the numerical iterative path integral summation 
method, from which we obtain observables of interest. We give a 
detailed discussion of the convergence properties of our method 
 and describe the extrapolation 
scheme.  
For several sets of parameters, we will present  
results and benchmark checks in Sec.\ \ref{sec:Results}. In addition, results for the mixed valence regime of the Anderson 
model are presented. A good agreement with tDMRG and fRG is found.
The Anderson Holstein model is studied by means of the ISPI method in 
Sec.~\ref{AHmodel}, the main finding of a sustained Franck-Condon blockade at low temperatures is discussed upon 
validating our spin-1 mapping of the AH Hamiltonian. By far the most complex model of this work, a magnetic interacting 
QD is studied in Sec.\ref{magAnderson}. We briefly review the necessary changes in the appearing Keldysh generating function and discuss our results on the relaxation time and impurity polarization as well as the tunneling current. A summary is given in Sec.~\ref{conclusion}.

\section{Keldysh generating function for non-interacting impurity models}
\label{sec:Model}
We consider the generic Hamiltonian for a quantum dot that is coupled to 
metallic leads to the left and right side ($\hbar=1, k_B=1$)
\begin{eqnarray}
\mathcal{H}&=&H_{dot}+H_{leads}+H_T \nonumber\\
&=&\sum_\sigma E_{0\sigma} n_\sigma
+\sum_{kp\sigma}(\epsilon_{kp}-\mu_p)c^\dag_{kp\sigma}c_{kp\sigma}\nonumber\\
&&- \sum_{kp\sigma} \left[t_p c_{kp\sigma}^\dagger d_\sigma + h.c.\right].
\label{andham}
\end{eqnarray}
Here, $E_{0\sigma}=E_0+ \sigma B$ with $\sigma=\uparrow,
\downarrow = \pm$ is the  energy of a single electron with
spin $\sigma$ on the isolated 
dot.  Tuning a back gate
voltage or a Zeeman magnetic field term $\propto B$ changes the value of $E_{0\sigma}$. The latter is 
assumed not to affect the electron dispersion in the leads. 
The corresponding dot electron annihilation/creation operator is 
$d_\sigma/d_\sigma^\dagger$, with  the density operator $n_{\sigma}\equiv d^\dag_\sigma
d_\sigma$. Possible eigenvalues of $n_\sigma$ are $\nu=0,1$, corresponding to the empty or occupied electronic state with spin $\sigma$. 
Interactions on the quantum dot are treated in Sec.~\ref{HStrafo} and not considered for the moment. 
In  Eq.\ (\ref{andham}), $\epsilon_{kp}$ denotes 
the energies of the noninteracting 
electrons (operators $c_{kp\sigma}$) 
in lead $p=L/R=\pm$, with chemical potential $\mu_p=p eV/2$. 
Quantum dot and leads are connected by the tunnel couplings $t_p$. 
The observable of interest is the (symmetrized) tunneling 
current $I=(I_L-I_R)/2$,
\begin{equation}
I(t)=-\frac{ie}{2}\sum_{kp\sigma}\left[pt_p\langle d^\dag_\sigma 
c_{kp\sigma}\rangle_t
-pt_p^*\langle c_{kp\sigma}^\dag d_\sigma\rangle_t\right],
\end{equation}
where $I_p(t) = -e \dot{N}_p(t)$ with $N_p(t)= 
\langle \sum_{k\sigma}c_{kp\sigma}^\dagger 
c_{kp\sigma} \rangle_t$. The stationary steady-state dc current follows as the
asymptotic long-time limit, $I=\lim_{t\to\infty}I(t)$. 
We have explicitly confirmed that current conservation, $I_L+I_R=0$, 
is numerically fulfilled  for the ISPI scheme.

In the presence of a finite bias voltage, $V\neq 0$, the Keldysh 
technique \cite{Keldysh,Rammer,Kamenev} provides a way 
to study nonequilibrium transport. In this formalism, 
the time axis is extended to a contour with $\alpha=\pm$ branches, see Ref.~\cite{Kamenev},
along with an effective doubling of fields. The Keldysh Green function (GF) 
exhibits a matrix structure 
$G^{\alpha\beta}_{ij}(t_\alpha,t'_\beta)=-i\langle \mathcal{T}_C[\psi_i(t_\alpha)
\psi^\dag_j(t'_\beta)]\rangle,$
where $\mathcal{T}_C$ denotes the contour ordering of times along the 
Keldysh contour, 
and $i,j=L,R,0$ correspond to  fields 
representing lead or dot fermions, respectively. 
We omit the spin indices here, remembering that 
each entry still is a diagonal $2\times 2$ matrix in spin space. 
The Keldysh partition function
contains all relevant information about the physics of the system. 
In order to obtain it, we first  integrate 
over the noninteracting lead fermion fields.  
Subsequently, we integrate over the dot fields as well.  
 In a fermion coherent state basis, the generating function is 
\begin{equation} \label{genfunc}
Z[\eta]=\int \mathcal{D}\left[\prod_\sigma \bar{d}_\sigma , d_\sigma ,
\bar{c}_{kp\sigma} ,
c_{kp\sigma}\right]e^{iS[\bar{d}_\sigma, d_\sigma ,
\bar{c}_{kp\sigma}, c_{kp\sigma}]},
\end{equation} 
with Grassmann fields  $(\bar{d}_\sigma,d_\sigma,\bar{c},c)$. The 
 external source term, which allows to compute the 
current at measurement time $t_m$, is chosen such that 
\begin{equation} \label{gencurrent}
I(t_m)=\left.-i\frac{\partial}{\partial\eta}\ln Z[\eta]\right|_{\eta=0}.
\end{equation}
Correspondingly, it is also possible to evaluate other 
observables, e.g., the zero-frequency shot noise, by introducing appropriate source
terms and performing the corresponding derivatives. 
The action is $S=S_{dot}+S_{leads}+S_T+ S_{\eta}$, see Ref.~\cite{ispi} for the explicit expressions. 
After integrating over the leads' degrees of freedom,   
the effective action for the dot becomes nonlocal 
in time. The generating function for the noninteracting system reads
\begin{eqnarray}
Z_{ni}[\eta]=\int \mathcal{D}\left[\prod_\sigma \bar{d}_\sigma d_\sigma
\right]e^{i(S_{dot,0}+S_{env})}
\end{eqnarray}
with
\begin{eqnarray}
S_{env}&=& \int_Cdt\int_C dt' \sum_\sigma\bar{d}_\sigma(t)
\bigg\{\gamma_L(t,t')+\gamma_R(t,t')\nonumber\\
&+&\frac{ie\eta}{2}[\gamma_L(t,t')-\gamma_R(t,t')] \nonumber \\
&\times & 
[\delta(t-t_m)+\delta(t'-t_m)]\bigg\}d_\sigma(t').
\label{Seff} 
\end{eqnarray}
For the source term, the physical measurement time $t_m$ is  fixed on the upper $(+)$ branch, hence the $(--)$ Keldysh 
element of the source term self-energy vanishes.
The $\gamma_p(t,t')$ matrices in Eq.\ (\ref{Seff}) 
represent the  leads, their Fourier transforms in frequency space are explicitly 
given as $2\times 2$ Keldysh matrices  
\begin{equation}
\gamma_p(\omega)=i\Gamma_p\left(
\begin{array}{cc}
2f(\omega-\mu_p)-1&-2f(\omega-\mu_p)\\
2-2f(\omega-\mu_p)& 2f(\omega-\mu_p)-1
\end{array}\right). 
\end{equation}
With the usual assumption that the 
 leads are in thermal equilibrium,  $f(\omega)=1/(e^{\omega/T}+1)$. 
Taking the wide-band limit  with 
a constant density of states $\rho(\epsilon_F)$ per spin channel around the
Fermi energy, the 
hybridization $\Gamma_p=\pi\rho(\epsilon_F)|t_p|^2$ of the dot level
with lead $p$ enters. We focus on symmetric contacts 
 $\Gamma_L=\Gamma_R\equiv \Gamma/2$ and on symmetrically applied bias voltages
as well.
The generalization to asymmetric contacts is straightforward. 

In  the next step (still for vanishing on-dot interactions), we integrate over
the dot degrees of freedom. This yields the noninteracting generating function 
\begin{equation}
Z_{ni}[\eta]= \prod_\sigma \det \left[
i G_{0\sigma}^{-1}(t,t')+\eta \Sigma^J(t,t') 
\right].
\end{equation}
The function $G_{0\sigma}^{-1}(t,t')$ follows from 
\begin{eqnarray}\label{fullD}
G_{0\sigma}(\omega) & = & 
\left[(\omega-\epsilon_{0\sigma})\tau_z-\gamma_L(\omega)-\gamma_R(\omega)\right]^{-1}
\nonumber \\
& = & \frac{1}{1+[(\omega-\epsilon_{0\sigma})/\Gamma]^2}\nonumber\\
&\times&\left(
\begin{array}{cc}
\omega-\epsilon_{0\sigma}+i\Gamma(1-F)&i\Gamma F\\
i\Gamma(F-2)&-\omega+\epsilon_{0\sigma}+i\Gamma(1-F)
\end{array}
\right),\, \nonumber
\end{eqnarray}
 where  
$\tau_z$ is the standard Pauli matrix in Keldysh space, 
and $F=f(\omega+eV/2)+f(\omega-eV/2)$.  Moreover, the self-energy 
for the source term is obtained as
\begin{eqnarray}
\Sigma^J (t, t') &=& \frac{e}{2} \left[\gamma_L(t, t') - 
\gamma_R(t, t')\right] \nonumber \\\label{sigmaJ}
& \times & \left[ \delta(t-t_m)+\delta(t'-t_m)\right] \, .
\end{eqnarray}
Up to this point, we have discussed the noninteracting case. 
The next section describes how to include the electron-electron interaction.

\subsection{Hubbard-Stratonovich transformation}
\label{HStrafo}
In the presence of on-dot Coulomb interactions, we add the Coulomb term 
\begin{equation}
\label{HU}
H_{int}=Un_{\uparrow}n_{\downarrow} \, 
\end{equation}
to the
Hamiltonian in Eq.~\eqref{andham}. 
For our purpose,  it is  convenient to use the operator identity 
$n_\uparrow n_\downarrow = \frac{1}{2}(n_\uparrow + n_\downarrow)
-\frac{1}{2}(n_\uparrow - n_\downarrow)^2 $, which results  in a shift of
the single-particle energies $\epsilon_{0\sigma}\equiv E_{0\sigma}+U/2$, and we
may rewrite  
$H_{dot}=H_{dot,0}+H_U=
\sum_\sigma \epsilon_{0\sigma} n_\sigma 
- \frac{U}{2}(n_\uparrow-n_\downarrow)^2 $. 
Now, the action in Eq.\ (\ref{genfunc}) in real time contains quartic terms of
dot Grassmann fields  and a Gaussian 
integration is not possible. We will use a time-discrete path integral
for the following discussion \cite{negele}. 
In order to decouple the quartic term, we 
discretize the full time interval, $t = N \delta_t$, with the 
time increment $\delta_t$. On each time slice, we perform a Trotter 
breakup of the dot propagator according to 
$e^{i\delta_t(H_{0}+H_T )} = e^{i\delta_t H_T/2}e^{i\delta_t H_{0}}
e^{i\delta_t H_T/2}+O(\delta_t^2)$,
where $H_0=H_{dot}+H_{leads}$. According to Refs.~\cite{HirschFye,Hirsch,Fye} 
the emerging Trotter error can be systematically 
eliminated from the results \cite{deRaedt,Eckel}, see below. 
On a single Trotter slice, a discrete Hubbard-Stratonovich transformation
\cite{Hirsch,Hubbard,Siano,roland} 
allows to decouple the interaction.This locally introduces 
Ising-like discrete spin fields $s_n =(s^+_n, s^-_n)$ 
on the $\alpha=\pm$ branches of the 
Keldysh contour with $s^\alpha_n = \pm 1$ on  the $n$-th Trotter slice. 
For a given Trotter slice, we define 
\begin{equation}
e^{\pm i\delta_tU(n_\uparrow-n_\downarrow)^2/2}=
\frac{1}{2}\sum_{s^\pm=\pm}e^{-\delta_t\lambda_\pm s^\pm(n_\uparrow-n_\downarrow)}.
\end{equation}
The HS parameter is obtained from the equation, 
\begin{displaymath}
\cosh(\delta_t\lambda_\pm)=\cos(\delta_tU/2)\pm i\sin(\delta_tU/2),
\end{displaymath}
under the condition that $U>0$, see Ref.~\cite{ispi}.
Note that the arbitrarily chosen overall sign of $\lambda_\pm$ does not 
influence the physical result. Uniqueness of  this 
HS transformation requires  $U\delta_t<\pi$.
To ensure sufficiently small time discretizations,  we meet the condition
${\rm max}(U, e|V|, |\epsilon_0|, T) \lesssim 1/\delta_t$ in all
calculations in general.

After the HS transformation, 
the remaining fermionic Grassmann variables $(\bar{d}_\sigma,d_\sigma)$  
appear quadratically and are integrated out at the cost of the full path
summation over the discrete HS Ising spins $\{s\}$ according to 
\begin{equation}
Z[\eta]=\sum_{\{s\}}\prod_\sigma \det G_\sigma^{-1}[\{ s \},\eta]  \, .
\label{pathsum}
\end{equation}
The full  Keldysh GF written in time-discretized 
($1\le k,l \le N$) form is
\begin{equation}
\left(G_{\sigma}^{-1}\right)_{kl}^{\alpha \beta}[\{ s \},\eta] 
= \left(G^{-1}_{0\sigma}\right)_{kl}^{\alpha \beta}
+i\eta\Sigma^{J,\alpha \beta}_{kl}
-i\delta_t \delta_{kl}\lambda_\alpha s^\alpha_k
\delta_{\alpha \beta},
\end{equation}
where $\alpha,\beta=\pm$ labels the Keldysh branches,
and the noninteracting GF is 
\begin{equation}\label{gij}
 G_{0\sigma,kl}=\int_{-\infty}^{\infty}\frac{d\omega}{2\pi}
e^{i\delta_t(k-l)\omega} G_{0\sigma}(\omega)  .
\end{equation}
Note that $G_{0\sigma}(t,t')$ depends only on time differences 
due to time-translational invariance of the noninteracting part which holds at
thermal equilibrium of the leads.  
The respective time discrete version of the self-energy kernel for the source
term is, cf. Eqs.~\eqref{Seff} and \eqref{sigmaJ},
\begin{equation}
\Sigma^{J,\alpha\beta}_{kl}=\frac{e}{2 }
\left[ \gamma_{L,kl}^{\alpha\beta}- \gamma_{R,kl}^{\alpha\beta}\right]
\frac{ \delta_{mk} \delta_{\alpha,+}+\delta_{ml} \delta_{\beta,+}}{\delta_t},
\end{equation} 
where $\gamma_{p,kl}=\gamma_p(t_k-t_l)$ and the measurement time
is $t_m=m \delta_t$.

\section{Formulation of the iterative scheme}
\label{sec:ISPI}
\begin{figure}
\centering
\includegraphics[width=\columnwidth]{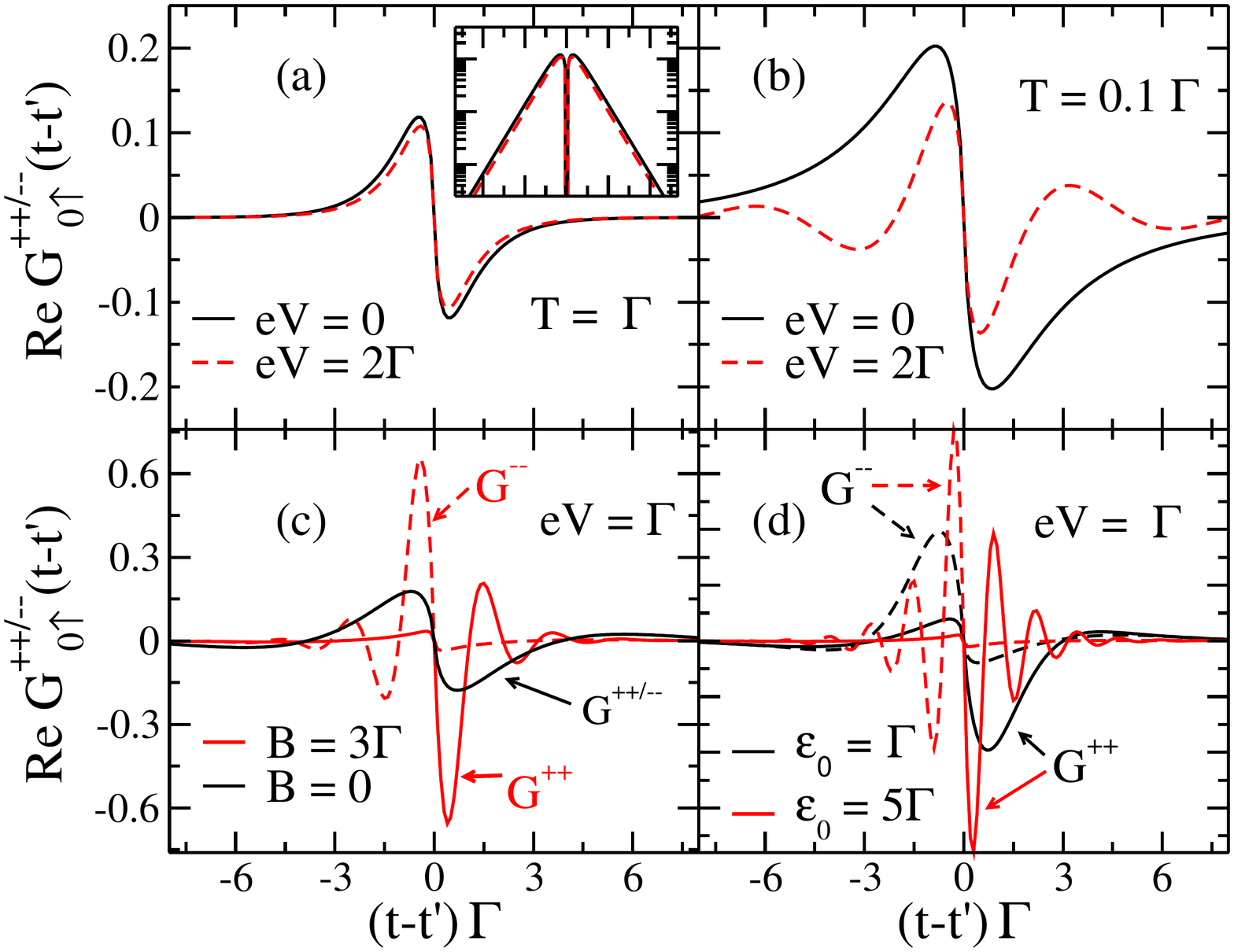}
\caption{(Color online) \label{corrfun}
Green's function of the non-interacting quantum dot in the presence of the leads for different settings of 
gate and bias voltages as well as magnetic field. The inset in (a) shows 
 $|{\rm Re \,} G^{--}_{0\uparrow}(t-t')|$ in a log-linear representation. }
\end{figure}
In this section we review the construction for the iterative expressions for the Keldysh partition
function which form the ISPI scheme. 
For this, we exploit the property (see, e.g., Ref.\
\cite{ispi,roland,DanielNJP,njp}) that each Keldysh 
component of $G_{0\sigma}(t-t')$ in Eq.~\eqref{gij} decays exponentially 
at long time differences ($\delta_t|k-l|\to\infty$) for any finite temperature, see Eq.\
(\ref{gij}).  The related   
time scale is denoted as correlation or memory time $\tau_c$.  
In Fig.\ \ref{corrfun}, we show typical examples of 
${\rm Re \,} G^{--}_{0\uparrow}(t-t')$ for different  bias 
voltages $V$. The exponential decrease with time 
is presented in the inset of Fig.\ \ref{corrfun}(a),  where 
the absolute value $|{\rm Re \,} G^{--}_{0\uparrow}(t-t')|$ is plotted in 
log-linear representation for the same 
parameters. For large bias voltages
 and low enough temperatures, e.g., at  $eV\gtrsim \Gamma$ and 
 $T\leq 0.2 \Gamma$, the decay is superposed by an oscillatory behavior. Since
the  lead-induced  correlation function decays as 
$\sim \cos[eV (t-t')/2]/\sinh[\pi T(t-t')]$, the  
respective correlations decay on a time scale 
given by $\tau_c^{-1}\sim \max(k_B T,eV)$. Thus, the exponential decay
 suggests to neglect lead-induced correlations beyond $\tau_c$. This motivates
an iterative scheme which exactly accounts for correlations within an interval
$\tau_c$, while neglecting them outside. Notice that the exponential decay is only present for 
finite $T$ and/or $V$, whereas at  $T=V=0$ correlations die out only
algebraically, and our approach is not applicable. 

Let us then face the remaining path sum in Eq.\ (\ref{pathsum}). 
In the discrete time representation, we denote
by $t_0=0<t_N=N\delta_t$ the initial and final times and  $t_k=k\delta_t$. 
The discretized GF and self-energy kernels for given spin $\sigma$ 
 are then represented as matrices 
of dimension $2N\times 2N$. 
For explicit calculations, we arrange the matrix elements 
 related to Keldysh space (characterized by the Pauli matrices 
 $\tau$) and to  physical times $(t_k,t_l)$ as 
$\tau\otimes(k,l)$. 
In particular, the ordering of the matrix elements 
from left to right (and from top to bottom) 
follows increasing times. 
The lead-induced correlations thus decrease exponentially   
with growing distance from the diagonal of the matrix. 

For numerical convenience, we evaluate the generating  function in the 
equivalent form
\begin{equation}
Z[\eta]={\cal N}\sum_{\{s\}}\prod_\sigma \det D_\sigma[\{s\},\eta] ,
\label{pathcom}
\end{equation}
with $D_\sigma= G_\sigma^{-1}G_{0\sigma} $, see Ref.~\cite{ispi} for details,
and thus get 
\begin{equation}
D_{\sigma,kl}^{\alpha\beta}[\{ s \},\eta] =
 \delta_{\alpha \beta}\delta_{kl}+ i \delta_t\lambda_\alpha  G_{0\sigma,kl}^{\alpha\beta} 
s^\alpha_k -i\eta\sum_{j,\alpha'} G_{0\sigma, kj}^{\alpha\alpha'}
\Sigma^{J,\alpha'\beta}_{jl}\,. 
\label{dij}
\end{equation}
By construction, we have to sum over $2N$ auxiliary Ising spins,  which appear
line-wise.  
The total number of possible spin configurations is $2^{2N}$. 

Next, we exploit the truncation  of the GF  by setting 
 $D_{kl}\equiv 0$ for $|k-l|\delta_t >\tau_c$, where 
 $\tau_c \equiv K \delta_t$
is the correlation time, with the respective number $K$ of Trotter time slices. All GF matrices have a band structure whose
band width is given by $K$. Equivalently, we can use for the spin-spin correlation the definition
\begin{equation}
\label{spinspin}
s_i^\alpha\cdot s_j^\beta=\left\{
\begin{array}{cc}
s_i^\alpha\cdot s_j^\beta, & \mbox{if} |i-j|\leq K,\\
	0,& \mbox{else}.
\end{array}
\right.
\end{equation}
We note that in the continuum limit, i.e., $K= N, \delta_t\to 0$ and 
$N\to \infty$,  the approach is formally exact. 

To proceed, we exploit that the determinant of a quadratic block matrix 
$D=\left(\begin{array}{cc}
a&b\\c&d \end{array}
\right)$
is given by
$\det(D)=\det(a)\det(d-ca^{-1}b)$. 
\begin{widetext}
In time space, we obtain the $(N\times N)$-Keldysh GF band matrix
\psset{xunit=1cm, yunit=1cm, linestyle=solid}
\begin{displaymath} 
  D \equiv  D_{(1,N_K)}
   = \left(
    \begin{array}{cccccc}
      D^{11}&D^{12}&0&0 & \dots & 0\\
      &&&&&\\
      D^{21}&D^{22}&D^{23}&0&
      \dots& \vdots\\
      &&&&&\\
      0&D^{32}&D^{33}&D^{34}&
      \dots &\vdots\\
      &&&&&\\
      0&0&D^{43}&D^{44}& \dots & 0 \\
      \vdots & \vdots &  \vdots & \vdots & \ddots &
      D^{N_K-1 N_K}\\
      &&&&&\\
      0 & \dots & \dots & 0 & D^{N_K N_K-1} &
      D^{N_K N_K}\\
    \end{array}
  \right),
\end{displaymath}
where the single blocks are $(K\times K)-$block matrices defined as 
($l,l'=1, \dots, N_K$)
\begin{displaymath}
D^{ll'}=\left(
\begin{array}{cccc}
D_{(l-1)K+1,(l'-1)K+1}&\dots&D_{(l-1)K+1,l'K}\\
\vdots&\ddots&\vdots\\
D_{lK,(l'-1)K+1}&\dots&D_{lK,l'K}
\end{array}
\right)\, .
\end{displaymath}
Without loss of generality, the number $N$ of Trotter slices is  chosen as
integer such that $N_K\equiv N/K$ and 
the matrix elements $D_{kl}$ follow from Eq.\ (\ref{dij}). We keep their
dependence on the Ising spins $s_k^\pm$ implicit.   
Each $D_{kl}$ still has a $2\times 2-$Keldysh
structure and a $2\times 2$ spin structure. 
 If we apply the formula for the determinant from above, the generating function (\ref{pathcom}) is represented as 
\begin{eqnarray}
Z[\eta]&=&\sum_{s_1^\pm,\dots, s_N^\pm}\det\left\{D^{11}[s_1^\pm,\dots, s_K^\pm]\right\} 
\nonumber \\
& \times & \det\left\{D_{(2,N_K)}[s_{K+1}^\pm,\dots, s_N^\pm]
-D^{21}[s_{K+1}^\pm,\dots,s_{2K}^\pm] 
\left[ D^{11}[s_1^\pm,\dots,s_K^\pm]\right]^{-1}
D^{12}[s_{K+1}^\pm,\dots,s_{2K}^\pm]
\right\} \, ,
\label{zeta}
\end{eqnarray}
where the  $(N_K-1)\times (N_K-1)-$matrix $D_{(2,N_K)}$ is obtained from
$D_{(1,N_K)}$ by removing
the first line and the first column. 

In order to set up an iterative scheme, we use the following
observation: to be consistent with the
 truncation of the correlations after a memory time $K\delta_t$, 
 we have to neglect terms that 
directly couple Ising spins at time differences 
larger than $\tau_{c}$, see also Eq.~\eqref{spinspin}, consequently matrix products of the form
\begin{equation}
D^{l+2,l+1}\left[D^{l+1,l+1}\right]^{-1}D^{l+1,l}\left[D^{l,l}\right]^{-1} D^{l,l+1}\left[D^{l+1,l+1}\right]^{-1}D^{l+1,l+2}=0
\end{equation}
within the Schur complement in each further iteration step. We do not neglect the full
Schur complement but only those parts which are generated in the
second-next iteration step. 
With this, we rewrite the generating function  as 
\begin{eqnarray} \label{zeta2}
Z[\eta] & = & \sum_{s_1^\pm,\dots, s_N^\pm} 
\det \left\{ D^{11}[s_1^\pm,\dots,s_K^\pm]\right\}
\prod_{l=1}^{N_K-1}\det \Big\{ D^{l+1,l+1}[s_{lK+1}^\pm,\dots,s_{(l+1)K}^\pm]
 \nonumber \\
 && -D^{l+1,l}[s_{lK+1}^\pm,\dots,s_{(l+1)K}^\pm]
\left[D^{l,l}[s_{(l-1)K+1}^\pm,\dots,s_{lK}^\pm]\right]^{-1}
D^{l,l+1}[s_{lK+1}^\pm,\dots,s_{(l+1)K}^\pm]
\Big\} \, . 
\end{eqnarray}
Exchanging the sum and the product, and reordering the sum  over all Ising 
spins, we obtain
\begin{equation}
Z[\eta] = \sum_{s_{N-K+1}^\pm, \dots, s_N^\pm} Z_{N_K} 
[s_{N-K+1}^\pm, \dots, s_N^\pm] \, ,
\end{equation}
where $Z_{N_K}$ is the last element obtained from the 
iterative procedure defined by ($l=1, \dots, N_K-1$)
\begin{equation}
Z_{l+1}[s_{lK+1}^\pm,\dots,s_{(l+1)K}^\pm]=
\sum_{s_{(l-1)K+1}^\pm,\dots,s_{lK}^\pm}
\Lambda_{l}[s_{(l-1)K+1}^\pm,\dots, s_{lK}^\pm,s_{lK+1}^\pm,\dots,s_{(l+1)K}^\pm]
Z_{l}[s_{(l-1)K+1}^\pm,\dots,s_{lK}^\pm] \, .
\label{zit}
\end{equation}
The  {\it propagating tensor} $\Lambda_{l}$ is read off
from Eq.\ (\ref{zeta2}) as
\begin{eqnarray}
\Lambda_{l}&=&\det \Big\{ D^{l+1,l+1}[s_{lK+1}^\pm,\dots,s_{(l+1)K}^\pm]
 \nonumber \\
 && -D^{l+1,l}[s_{lK+1}^\pm,\dots,s_{(l+1)K}^\pm]
\left[D^{l,l}[s_{(l-1)K+1}^\pm,\dots,s_{lK}^\pm]\right]^{-1}
D^{l,l+1}[s_{lK+1}^\pm,\dots,s_{(l+1)K}^\pm]
\Big\}\, .
\label{lambda}
\end{eqnarray}
The iteration starts with 
 $Z_{1}[s_{1}^\pm,\dots,s_{K}^\pm]=
 \det \left\{D^{11}[s_1^\pm,\dots,s_K^\pm]\right\}$. 
\end{widetext}

The current is numerically obtained by evaluating 
Eq.\ (\ref{gencurrent})  
for a small but fixed value of $\eta$; we have taken 
$\eta=0.001$ for all results shown. The current as a function of time
$I(t_m)$ shows a transient oscillatory or relaxation behavior at short times. 
We extract the stationary value $I$ when the current has saturated to a plateau.

\subsection{Convergence and extrapolation procedure}
\label{sec:extrapol}
\begin{figure}
\centering
\includegraphics[width=\columnwidth]{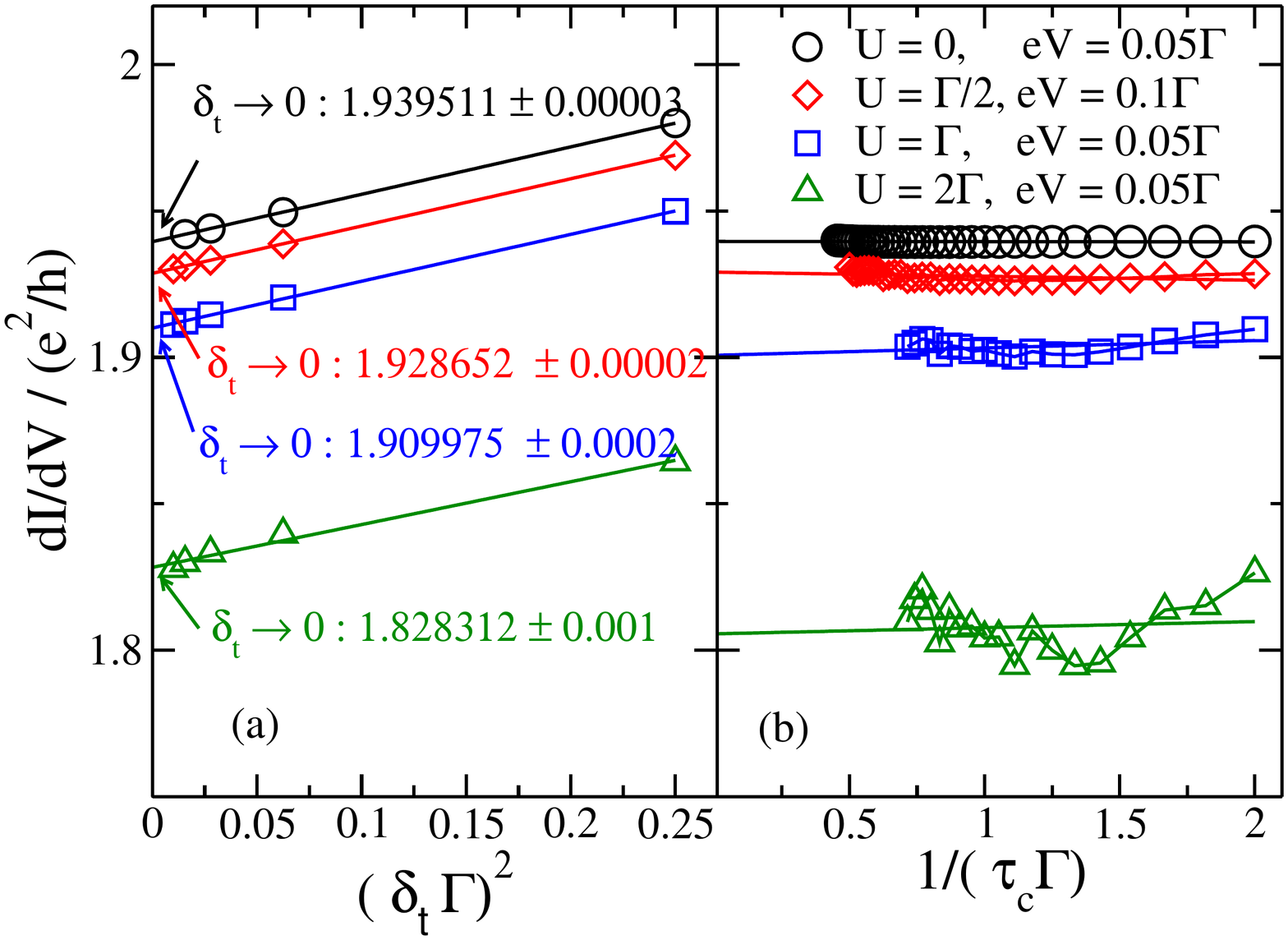}
\caption{(Color online) \label{tautrott}
Extrapolation scheme for raw data obtained with ISPI. First the Trotter error, panel (a) and then the memory 
truncation error (b) are extrapolated. Notice that whenever the ISPI data are converged, they are numerically exact. }
\end{figure}

By construction there are two systematic errors in the scheme: 
 (i) the Trotter error due to finite time
discretization $\delta_t=t/N$, and (ii) the memory error due to a
finite memory time $\tau_c = K \delta_t$. The scheme becomes exact 
in the limit $K\to \infty$ and $\delta_t \to 0$. 
We can eliminate both errors from the
numerical data in the following systematic way:  
Step 1: We choose a fixed
  time discretization $\delta_t$ and a 
  memory time $\tau_c$. A reasonable estimate for $\tau_c$ is 
 the minimum of $1/|eV|$ and $1/T$ (see
above). With that, we calculate the current $I(\delta_t, \tau_c)$, 
and, if desired, the differential conductance $dI(\delta_t, \tau_c)/dV$ 
(the derivative is performed numerically for
a small $\Delta eV=0.01 \Gamma$). The calculation is
then repeated for
different choices of $\delta_t$ and $\tau_c$.  
Step 2: Next, the Trotter error can be eliminated by exploiting 
the fact that it  vanishes quadratically for  
$\delta_t\to 0$ \cite{HirschFye,Hirsch,Fye}. For  a fixed  memory 
time $\tau_c$, we can thus extrapolate  and obtain 
$dI(\tau_c)/dV=dI(\delta_t \to 0, \tau_c)/dV$, which still
 depends on the finite memory time $\tau_c$. The quadratic dependence on
 $\delta_t$ is illustrated  in Fig.\  \ref{tautrott} (a) for different
 values of $U$. Note that each line corresponds to the same fixed
 memory time $\tau_c=0.5/\Gamma$.   
 Step 3: In a last step, we eliminate the memory error by  
 extrapolating to $1/\tau_c \to 0$, and obtain
 the final numerically exact value $dI/dV=dI(\tau_c \to \infty)/dV$. 
For the  dependence on $1/\tau_c$,  we empirically
 find a regular and systematic behavior as shown in Fig.\
 \ref{tautrott}(b). The $\tau_c \to \infty$ value is approached with
 corrections of the order of $1/\tau_c$, see Fig.\
 \ref{tautrott}(b). 
 
We have implemented the  iterative scheme together with the convergence
procedure on standard Xeon 2GHz machines. 
Computations are then only possible for $K\leq 7$ due to 
the limited memory resources available.
Typical running times for the shown simulation data are 
approximately $15$ hours for $K=5$. 

\section{Benchmarking the approach: 
comparison with exact and perturbative results}
\label{sec:Results}
\begin{figure}
\centering
\includegraphics[width=\columnwidth]{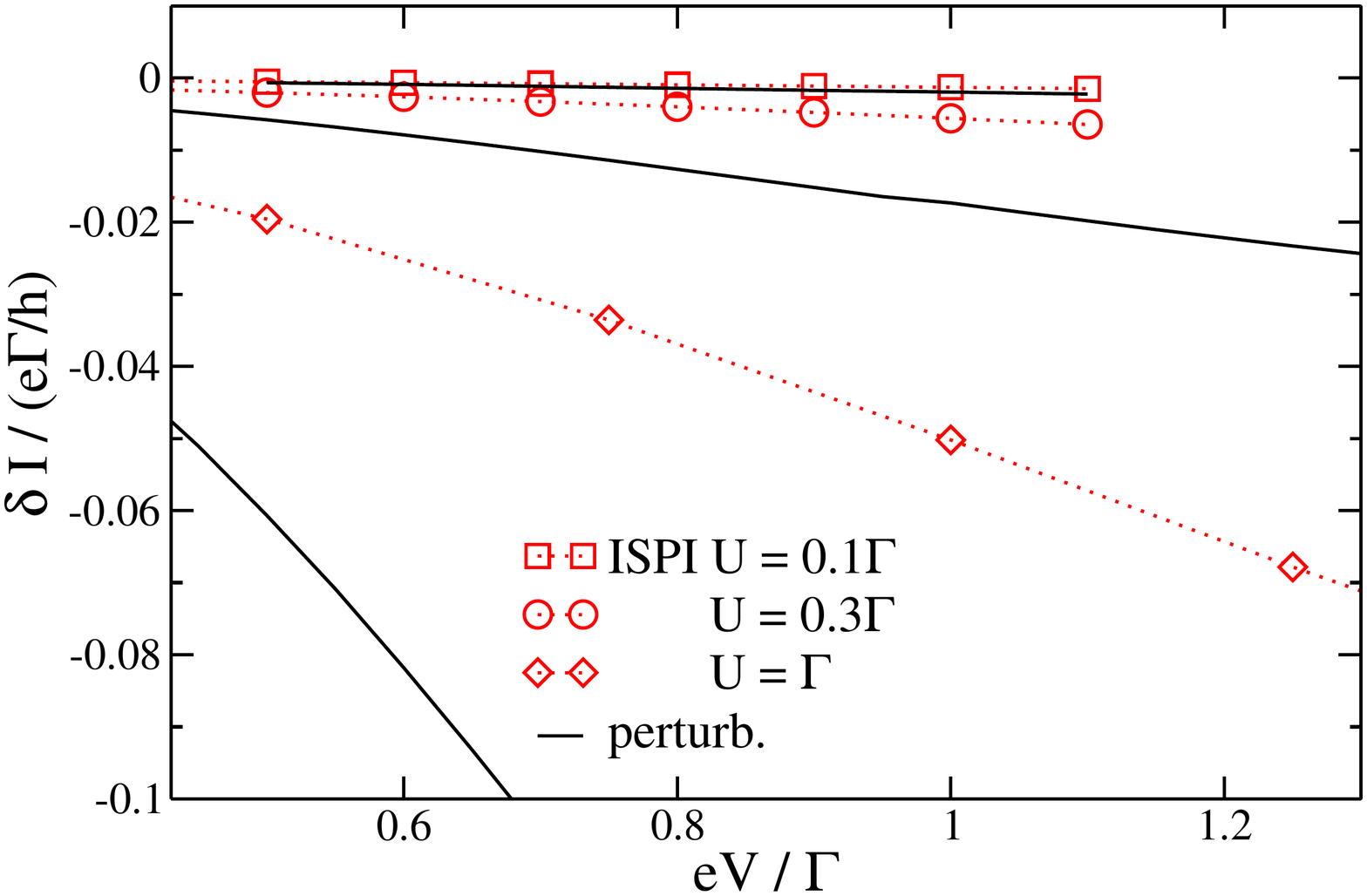}
\caption{(Color online) \label{pertu} Corrections $\delta I$ to the current due
to finite $U$ as compared to $U=0$ for small to intermediate on-dot interaction $U\leq \Gamma$. 
Other parameters are $T=0.1\Gamma$, $\epsilon_0=B=0$.
We compare the ISPI data (red symbols) to a 2nd order perturbative calculation, dotted lines guide the eyes.}
\end{figure}

In this section, we discuss the results obtained for 
the Anderson model. We measure energies in units of $\Gamma$. Unless noted
otherwise, all error bars for the shown data points, 
which are due to the Trotter and memory extrapolation scheme, are of 
the order of the symbol sizes in the figures.  

We stop short on reporting that we have recovered the exact current for the solvable $U=0$ case, 
as a function of the bias and the gate voltage.
By construction, the ISPI method includes all tunneling processes exactly to arbitrary orders.

In order to benchmark our code for finite $U$, we compare the numerical results 
to a perturbative calculation at the charge degeneracy point  $\epsilon_0=B=0$, 
where the interaction self-energy can be computed 
up to second order in $U$ \cite{Fujii,Luca}. For a detailed comparison, we plot 
mostly the {\sl interaction corrections},  $\delta A\equiv A(U)-A(U=0)$, 
with $A$ being the current $I$,  the linear conductance $G$,  or 
the nonlinear conductance $dI/dV$, respectively.    
Figure \ref{pertu} shows the results for $\delta I$ as a function 
of the bias voltage for $U=0.1 \Gamma$,  $U=0.3\Gamma$ and $U=\Gamma$. 
For $U=0.1 \Gamma$,  we perfectly recover the perturbative 
results, which confirms the reliability of our code even in the regime 
of nonlinear transport. Clearly, the corrections are small and negative, 
which can be rationalized in terms of indications of Coulomb blockade physics,
as transport is suppressed by a finite on-dot interaction. 
For $U=0.3\Gamma$, the current decreases even further, 
and the deviations between the ISPI and perturbative results increase.  
The relative deviation for $U=0.3\Gamma$ is already $\approx 30-35 \%$, 
illustrating that perturbation theory is already of limited accuracy in this 
regime. Although it well reproduces the overall tendency, there is a
significant quantitative difference. 
It is even more pronounced  for $U=\Gamma$, 
as shown in the figure. Here, second-order perturbation theory does not even reproduce qualitative 
features.

\subsection{Comparison with a master equation approach}
\begin{figure}
\centering
\includegraphics[width=\columnwidth]{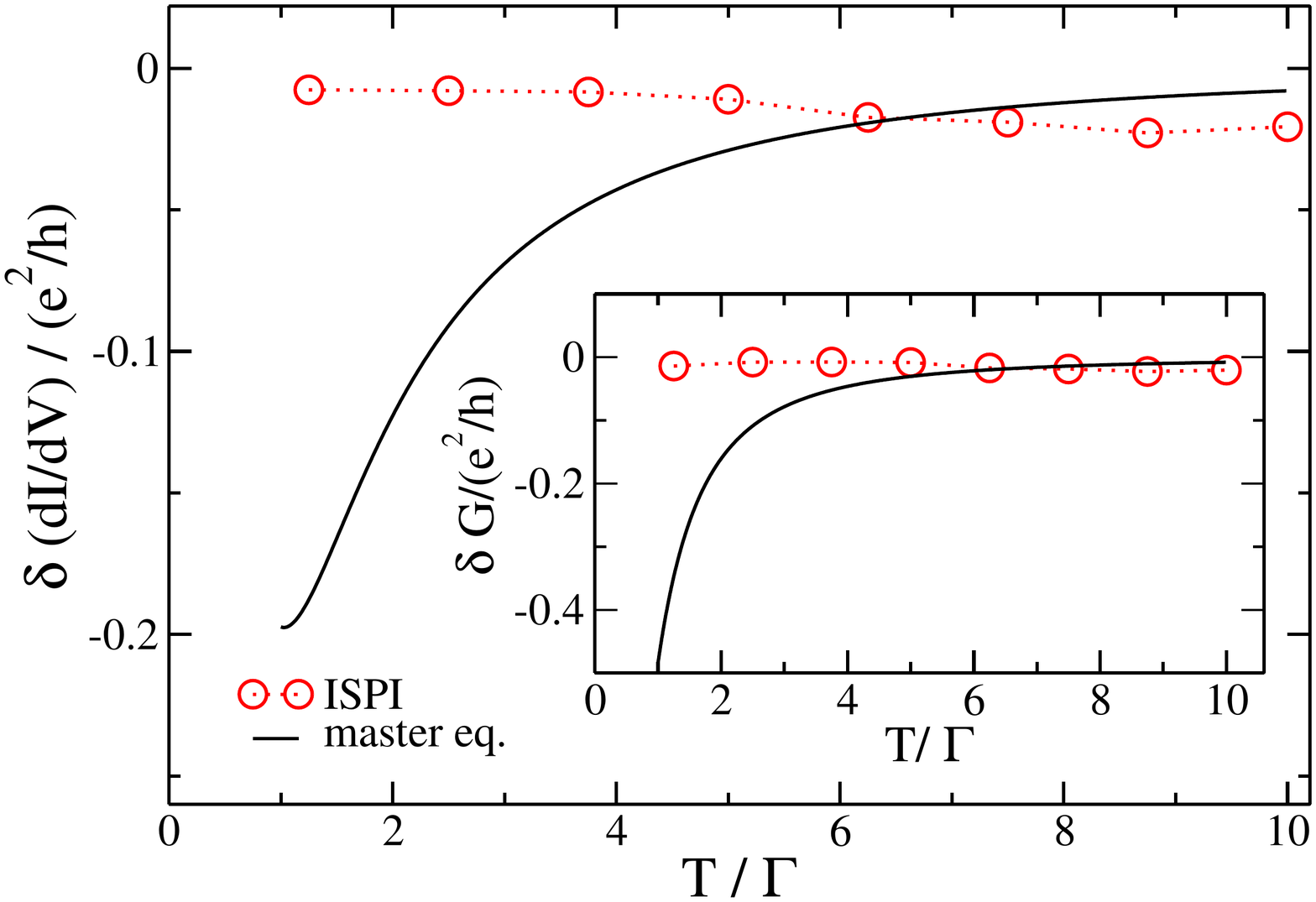}
\caption{(Color online) \label{masterlinnonlin}Interaction corrections for $U=\Gamma$ to the 
nonlinear (main) and linear conductance (inset) calculated by ISPI (symbols)
and by an incoherent rate equation as a function of temperature. In the main panel we have chosen 
$eV=3\Gamma$.}
\end{figure}
Next, we compare our numerically exact results with the outcome of  
a standard classical rate equation calculation \cite{Koenig,Flensberg}. 
The rate equation is expected to
yield reliable results in the incoherent (sequential) tunneling regime, when 
 $T \gg \Gamma$. Then, a description in terms of occupation probabilities
for the isolated many-body dot states is appropriate. 
Results for $U=\Gamma$ are shown in Fig.\ \ref{masterlinnonlin}, 
both for the interaction corrections to the nonlinear and the linear
 conductance. When the temperature is lowered, $T<\Gamma$, 
quantum coherent interaction effects become more important, as seen from the
exact ISPI results. They are clearly not captured by the master equation
in the sequential tunneling approximation.
  However, for $T\gg \Gamma$,
interaction corrections are washed out, and the master equation 
becomes accurate, cf. Fig.~\ref{masterlinnonlin}.  
Similarly, from our ISPI results, we have found (data not shown) that 
interaction corrections are suppressed by an increasing bias voltage as well.

\subsection{ISPI vs. tDMRG and fRG}
\begin{figure}
\centering
\includegraphics[width=\columnwidth]{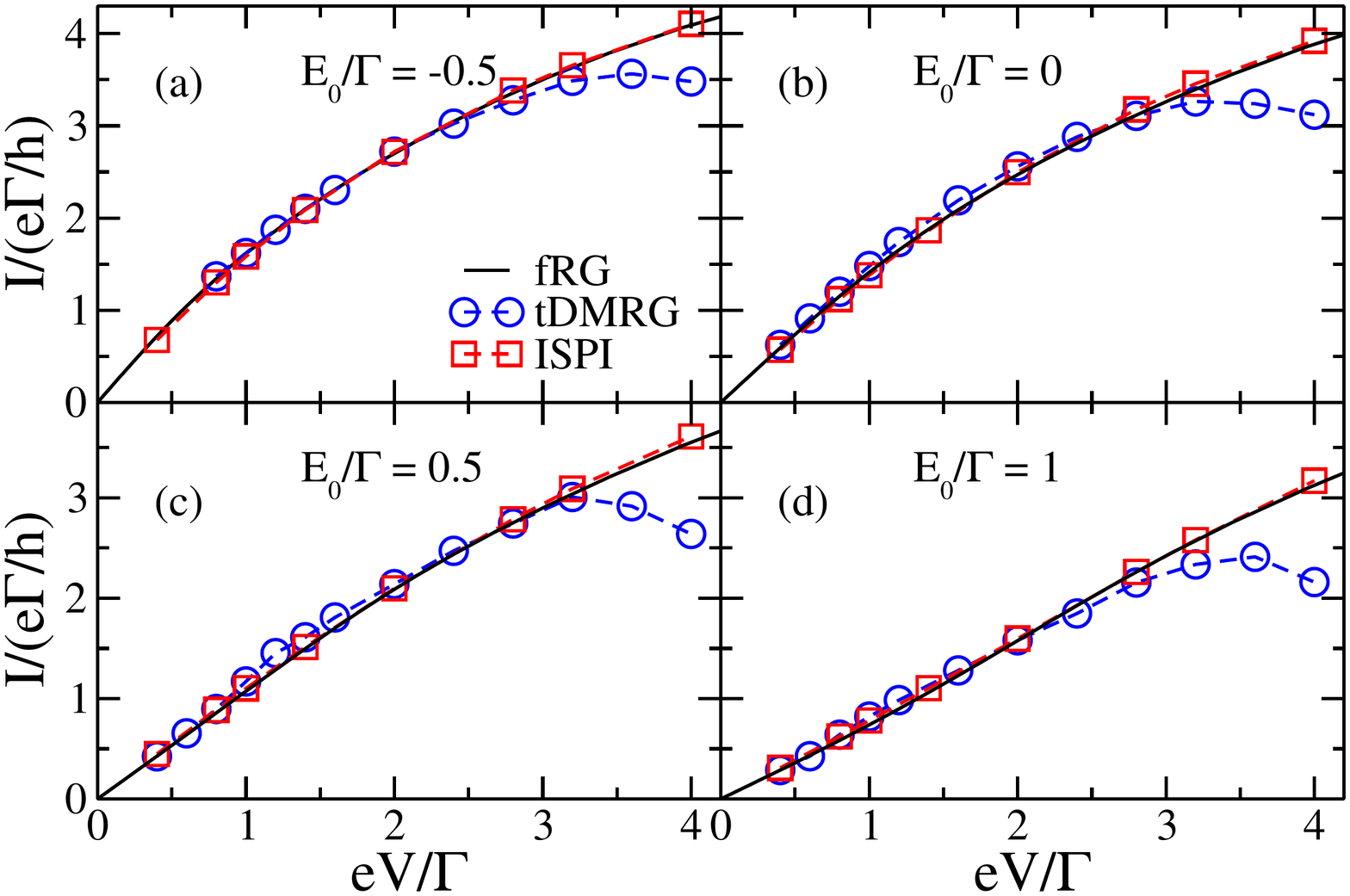}
\caption{(Color online) \label{compfrg} Comparison of ISPI and RG methods in the mixed valence regime. 
Parameters are chosen as $U/\Gamma=2$ and the temperature is $T= 0.1\Gamma$. The RG data are taken from Ref.~\cite{njp}}
\end{figure}
The mixed valence regime is characterized by the fact that the QD's occupation fluctuates between the different charge states, namely empty, singly and doubly occupied, see Ref.~\cite{Costi2} for details.  Furthermore the coupling to the leads is strong, i.e. the Kondo temperature $T_K\sim \Gamma$, see below in Sec.~\ref{sec:linear}. When controlling the gate voltage by tuning $\epsilon_0$, it is possible to tune the nanostructure into this limit. 
From the theoretical point of view another energy scale $\epsilon_0$ enters, again, not as a small parameter and simulation methods are rare in this regime. 
For the case of intermediate Coulomb repulsion $U/\Gamma=2$ we have studied \cite{njp} the mixed valence regime and compare our results 
to other state-of the art methods, functional renormalization group (fRG) \cite{Jakobs} and time dependent density matrix renormalization group (tDMRG)\cite{Schmitteckert,Schollwoeck}. The results are shown in Fig.~\ref{compfrg} for the steady state current $I(V)$ for $\epsilon_0\ne 0$, i.e., away from the charge degeneracy point.
The results from fRG and ISPI match perfectly from small bias voltages $eV / \Gamma \approx 0.2$ up to the strong non-equilibrium regime. We note that by construction, it becomes increasingly cumbersome (and finally impossible) to obtain converged ISPI results in the limit of vanishing bias voltages and low temperatures \cite{ispi}, since then the correlations do not decay sufficiently to be truncated. In the present setup, tDMRG has a tendency to overestimate the currents in the mixed valence regime, see Ref.~\cite{Heidrich} for details, hence we find slight deviations between tDMRG and the two other methods (see $eV \approx1.5\Gamma$, lower left panel). Overall agreement away from the symmetric point between the three methods is very good. Furthermore, $I(V)$ curves show agreement for a wide range of magnetic fields, applied to the QD, see Ref.~\cite{njp} for details.

\subsection{Small bias regime $eV \ll \Gamma$:}
\label{sec:linear}
\begin{figure}
\centering
\includegraphics[width=\columnwidth]{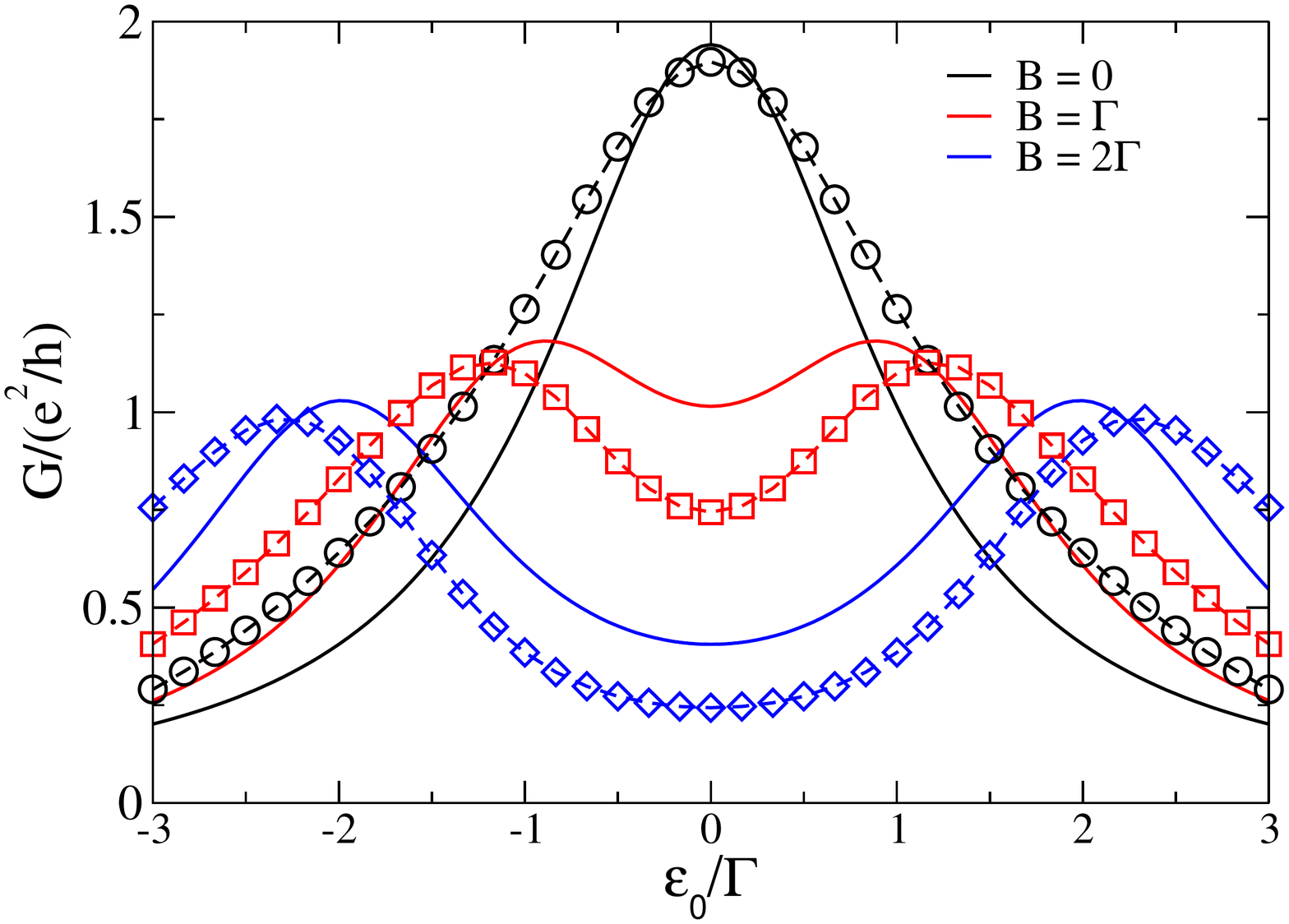}
\caption{(Color online) \label{Gvseps01}
Linear magneto-conductance as a function of the gate voltage $\epsilon_0$ and $U=\Gamma$. 
The bias voltage is chosen as $eV=0.05\Gamma$ and $T=0.1\Gamma$.
 ISPI data (symbols) are connected by dashed lines as guide to the eyes. The
full lines show the analytical results for vanishing interaction $U\to 0$. 
Finite magnetic fields split the resonant tunneling peak.}
\end{figure}

For sufficiently small bias voltage, 
the current is linear in $V$, and we can focus 
 on the linear conductance $G$. Figure \ref{Gvseps01} shows $G(\epsilon_0)$
for different magnetic fields $B$, 
taking $U=\Gamma$ and $T=0.1\Gamma$ (for $eV=0.05\Gamma$). 
For $B=0$, two spin-degenerate transport channels contribute,
and a single resonant-tunneling peak at $\epsilon_0=0$ results. 
For $B\ne 0$, the spin-dependent channels are split
 by $\Delta \epsilon_0=2B$, resulting in a double-peak
structure. Furthermore, since also $U$ lifts the degeneracy, the  spin-resolved
levels are now 
located at $\epsilon_0=\pm (B+U/2)$ due to the Zeeman splitting.
We find an interaction-induced broadening, cf.~Fig.\  \ref{Gvseps01}, 
of the resonant-tunneling peak as compared to the noninteracting case. 
The width of the Lorentzian peak profile for $B=0$ is determined by 
$\Gamma$ at sufficiently low $T$, and broadens as $T$ increases. 
Here, the double-peak structure, with two clearly separated
peaks for  finite $B$, is not yet fully developed. The two peaks largely 
overlap, and the distance of the peaks  is below the expected  
$\Delta \epsilon_0 =2B$, since tunneling  significantly 
broadens the dot levels. 
\begin{figure}
\centering
\includegraphics[width=\columnwidth]{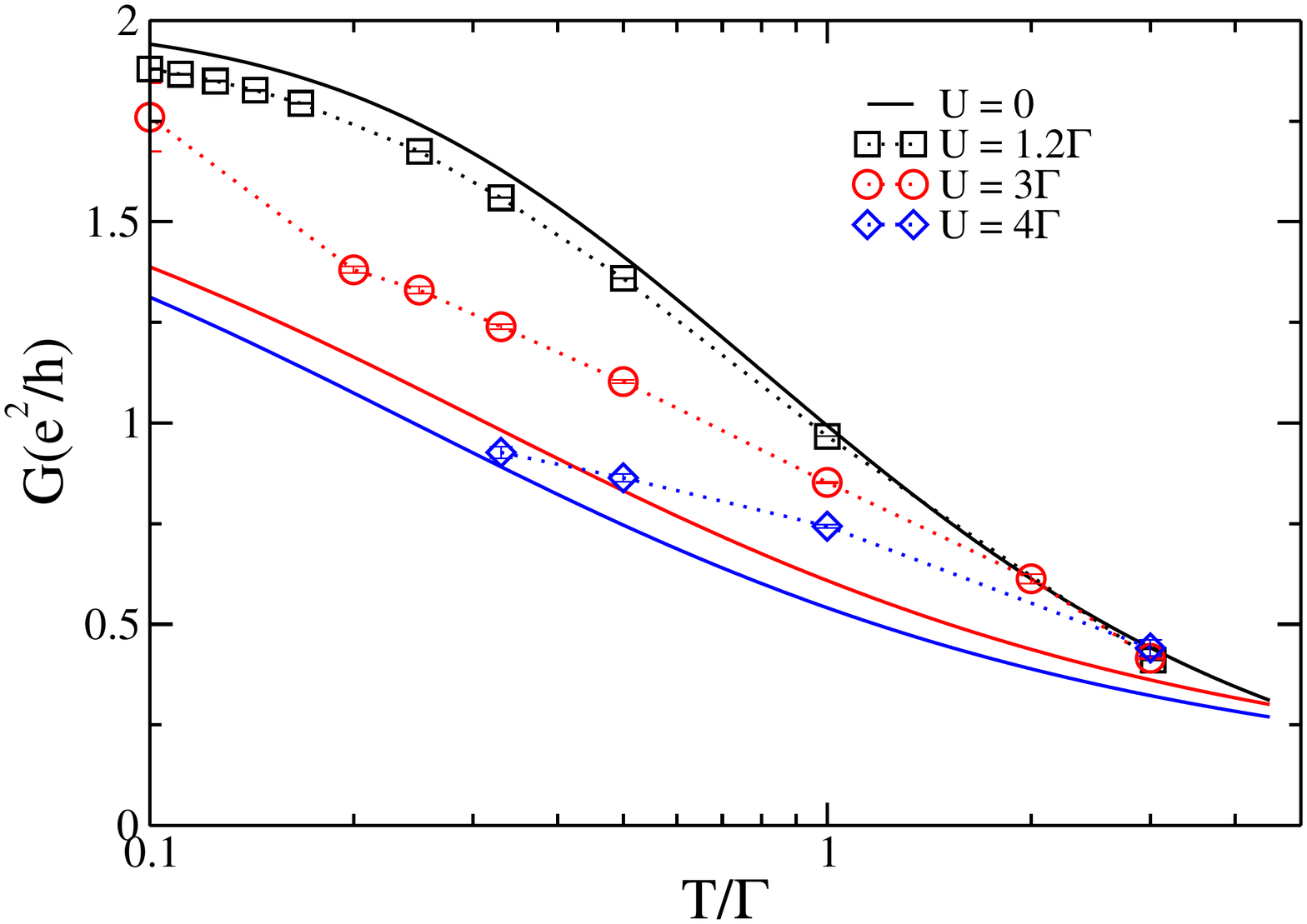}
\caption{(Color online) \label{lingoft} Log-linear plot of the linear conductance $G$ as a function 
of temperature. Above and close to the Kondo temperatures $T_{K}$, the ISPI
simulation results (symbols) agree with the results of Ref.\ \cite{Hamann}
given as full lines, see text.}
\end{figure}

Next, we address the temperature dependence of the linear conductance 
(numerically evaluated for $eV=0.05 \Gamma$). In Fig.\ 
\ref{lingoft}, we show $G(T)$ for different values of $U$ 
(up to $U=4 \Gamma$) at $\epsilon_0=B=0$. For $U=1.2 \Gamma$, the 
deviations from the $U=0$-result is small. 
For larger $U$, deviations become more pronounced at low temperatures
where the on-dot interaction becomes increasingly relevant.  
Up to present, we have obtained converged results in the regime 
of {\em small\/} bias voltages for interaction strengths 
$U\leq 4 \Gamma$ for temperatures above or close to the Kondo temperature,  
$T\gtrsim T_K$. 
The corresponding Kondo temperatures are $T_K=0.38\Gamma$ for 
$U=3\Gamma$ and $T_K=0.293 \Gamma$ for $U=4 \Gamma$. 
In the regime $T_K \lesssim T \lesssim 10 T_K$, we compare our results 
to the result of Hamann \cite{Costi2,Hamann},
\begin{equation}\label{hamann}
G(T)=\frac{e^2}{h} 
\left(1-\frac{\ln(T/T_{KH})}{[\ln^2(T/T_{KH})+3\pi^2/4]^{1/2}}\right) \, ,
\end{equation}
for the linear conductance, where $T_{KH}=T_K/1.2$, see Fig.\ \ref{lingoft} 
(solid lines). 
In Ref.\ \cite{Costi2}, it has been shown that the results of 
 the numerical RG coincide with those of Eq.\ (\ref{hamann}) in this regime. 
 Fig.\ \ref{lingoft} illustrates that 
 the agreement between the two approaches is satisfactory and 
 shows that the ISPI provides reliable results 
 in the linear regime above or close to the Kondo
 temperature. Due to the construction of the approach, the situation is 
more favorable for large bias voltages, where short to intermediate 
memory times allow us to obtain convergent results.

\subsection{Large bias regime $eV \ge \Gamma$:}
\label{sec:noneq-res}
\begin{figure}
\centering
\includegraphics[width=\columnwidth]{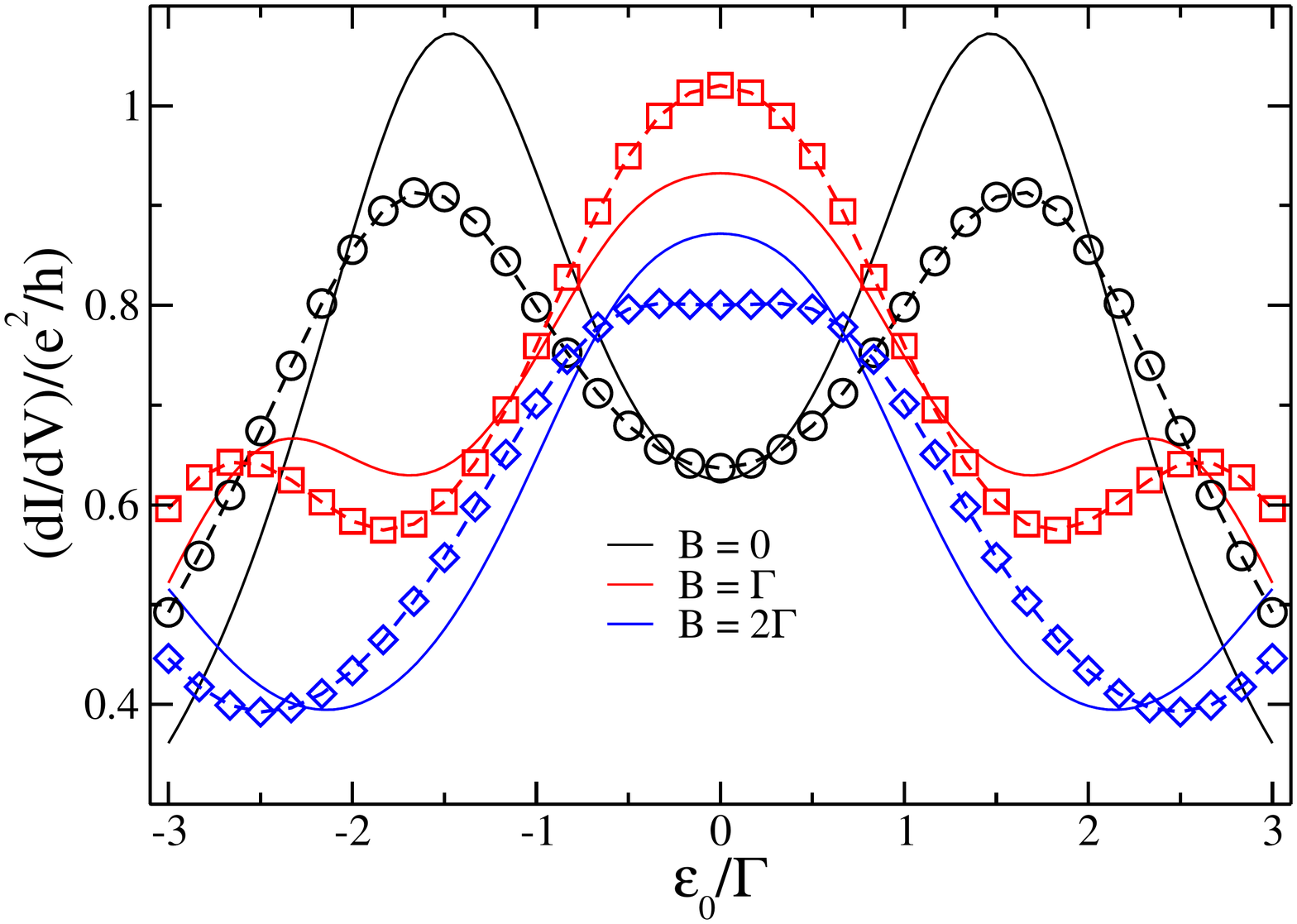}
\caption{(Color online) \label{Gvseps04}
Same as in Fig.~\ref{Gvseps01} but for a large bias voltage $eV=3\Gamma$.}
\end{figure}

Let us then turn to nonequilibrium transport for voltages  $eV\gtrsim \Gamma$.
Here,  the transport window is given by $\sim eV$, and a 
double-peak structure for $dI/dV$ emerges even for $B=0$, see Fig.~\ref{Gvseps04}
with  distance $eV$ between the peaks. We show results for $eV=3\Gamma$ but
otherwise the same parameters as in Fig.\ \ref{Gvseps01}.
 For an additional finite magnetic field, each
peak of the double-peak structure itself experiences an additional 
 Zeeman splitting, resulting in an overall four-peak structure. 
 For $B=\Gamma$ and the depicted values of $U=\Gamma$, 
 the two innermost peaks (closest to $\epsilon_0=0$) overlap
 so strongly that they effectively form a single peak at $\epsilon_0=0$ 
 again. The two outermost peaks are due to the combination of the 
 finite magnetic field and the bias voltage. 

Increasing the on-dot interaction $U$ downsizes 
the differential conductance peaks as compared to the noninteracting
case, i.e., the interaction corrections are again largest when the level 
energy matches the chemical potential in the leads. 
Note that the four-peak structure is already present in the
noninteracting case (with $B\ne 0$) 
and, hence, is not  modified qualitatively by a finite $U$. 

\begin{figure}
\centering
\includegraphics[width=\columnwidth]{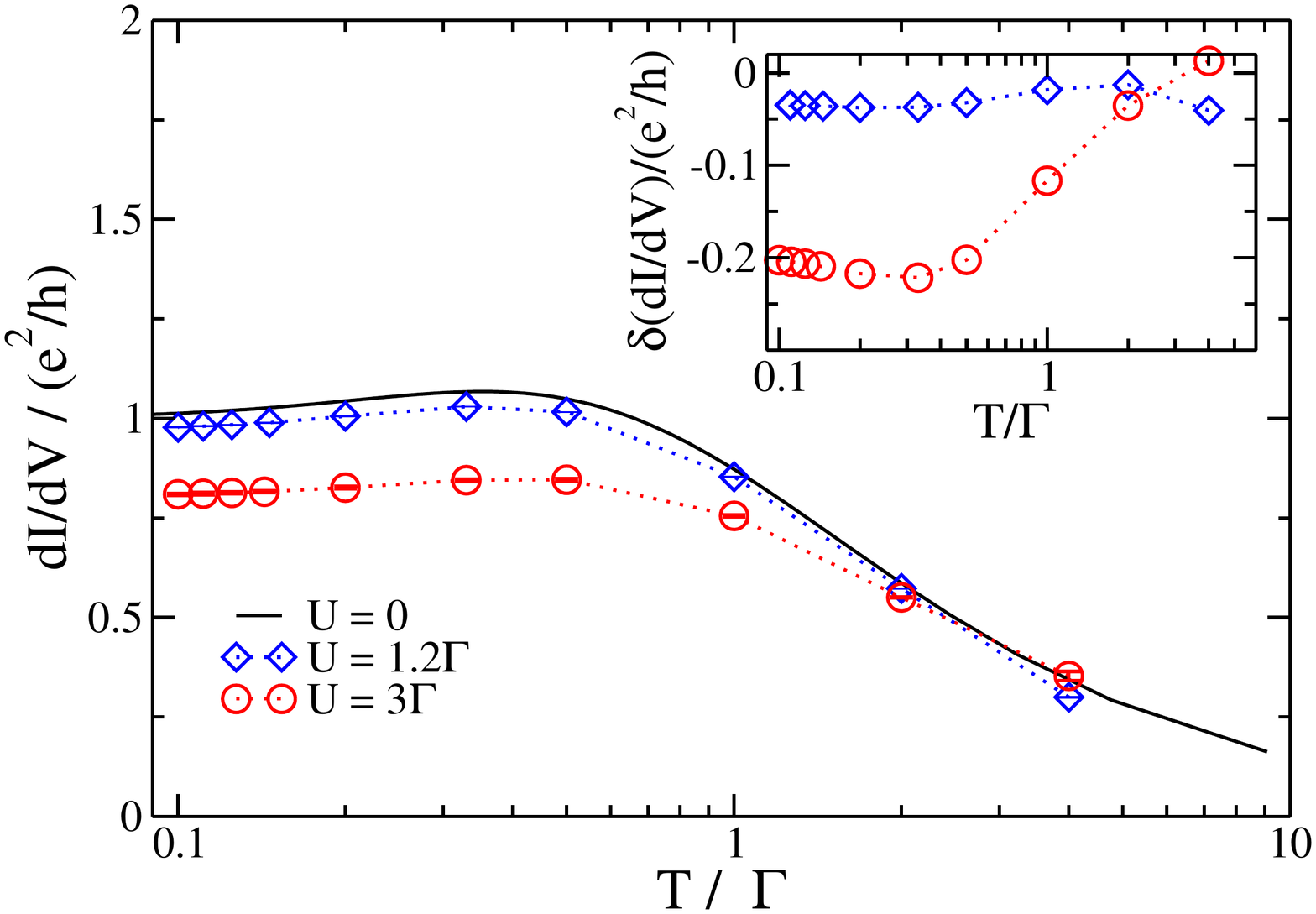}
\caption{(Color online) \label{GofT} Log-linear representation of the 
temperature dependence of the differential conductance 
for different interaction strengths $0\leq U/\Gamma<3$ for $eV=2 \Gamma,
\epsilon_0=B=0$. }
\end{figure}

Finally, we address the temperature dependence of 
the differential conductance $dI/dV$. 
ISPI results for $eV=2 \Gamma, \epsilon_0=B=0$ are shown in 
 Fig.\ \ref{GofT}. Again, as in the linear regime, 
 the conductance increases  with lower temperatures,  
and finally saturates, e.g.,  at $dI/dV=e^2/h$ for $U=0$ and $eV=2\Gamma$. 
Clearly the conductance decreases when the bias voltage is 
raised. 
Increasing $U$ renders this suppression yet more pronounced, see also 
inset of Fig.\ \ref{GofT} for the corresponding corrections. 
At high temperatures,  thermal fluctuations wash out the interaction 
effects, and the interaction corrections die out. 
\section{Sustained Franck-Condon blockade in molecular quantum dots}
\label{AHmodel}
As a second example, we apply the ISPI scheme to investigate vibrational
effects in the nonequilibrium tunneling current  through a molecular quantum
dot, see Ref.~\cite{roland}. We extend the single-impurity Anderson model by a linear phonon
of frequency $\Omega$ (annihilation operator $b$) which  
couples to a single spinless electronic level with energy $E_0$ (operators $d/d^\dag$). Hence, we
have the molecular Hamiltonian 
\begin{equation}\label{hm}
H_m =  \Omega \ b^\dagger b + \left[ E_{0}+\lambda (b+b^\dagger) 
\right ] n_d 
\end{equation}
with electron-phonon coupling strength $\lambda$ and $n_d=d^\dag d$.  

The short-time
propagator on the forward/backward branch of the Keldysh contour, 
$e^{\mp i\delta_t H}$, then allows for a Trotter breakup, 
$e^{\mp i\delta_t H} =e^{\mp i\delta_t H_1} e^{\mp i\delta_t(H-H_1)}$, 
with $H_1= H_m-\Omega b^\dagger b$, where the auxiliary relation
\begin{equation}\label{aux}
e^{\mp i \delta_t H_1}= 1 - n_d + n_d e^{-\lambda^2\delta_t^2/2}
e^{\mp i\delta_tE_0} e^{\mp i\delta_t\lambda b^\dagger} 
e^{\mp i\delta_t\lambda b}
\end{equation}
holds. 
This effectively decouples the electron-phonon interaction in terms of a 
three-state variable $s_{\eta}=0,\pm 1$ defined at each (discretized) time 
step $t_{j}$ along the forward/backward ($\alpha=\pm$) part of the 
Keldysh contour, where $\eta=(t_j,\alpha)$.
Below, we also use the notation $\eta\pm 1=(t_{j\pm 1},\alpha)$
with periodic boundary conditions on the Keldysh contour.
The ``Ising spin'' variable $s_\eta$ picks up the three terms in
Eq.~\eqref{aux} 
and acts like a Hubbard-Stratonovich auxiliary field, similar to 
the Ising field employed in the Hirsch-Fye formulation of
the Anderson model \cite{ispi,HirschFye,Hirsch,Fye}. The bosonic (phonon)
scalar field and the fermionic (dot and lead electrons) Grassmann fields
appearing in the Keldysh path integral are noninteracting but
couple to the time-dependent auxiliary spin variable.
Hence, those fields can be integrated out analytically and 
the time-dependent current $I(t_m)$ follows from a path summation as above in
Sec.~\ref{sec:Model}. The resulting 
matrix $D_{\eta\eta'}$ (in time and Keldysh space) depends on the 
complete spin path $\{s \}$. Specifically, we obtain $D=-iB(G_d^{-1}-\Sigma)$,
where $G_d^{-1}$ 
has spin-dependent matrix elements 
$\left[-iG^{-1}_d \right]_{\eta+1,\eta}=-s_{\eta}$.
We find $\Sigma_{\eta\eta'}\ne 0$ only when $s_{\eta}=\pm 1$,
where it coincides with the usual (wide-band limit) expression \cite{nazarov}. 
Finally, the diagonal matrix $B$ (quoted here for $\epsilon_{0}=0$) with
\begin{equation}
B_{\eta\eta}=A_{s_{\eta}} e^{-\lambda^2\delta_t^2 \sum_{\eta'} \alpha \alpha'
[iG_{ph}]_{\eta,\eta'+1} | s_\eta s_{\eta'} |}
\end{equation}
encapsulates all phonon effects, where
$G_{ph}$ is the discretized phonon Green's function, 
see Ref.~\cite{kamenev}, and we have used the notation
$A_{0}=1$ and $A_{\pm 1}=\pm (1/2) e^{-\lambda^2\delta_t^2/2}$.
We comment shortly on the peculiar convergence properties of the present model.
Convergence of the extrapolation 
requires intermediate $T$ or $V$-values, for otherwise the necessary memory 
times become exceedingly long.  For the results below, 
we have used $K\le 4$ and $0.3\le\Gamma\delta_t\le 0.35$.  The shown current
follows by averaging over the $\delta_t$-window, with error bars indicating 
the mean variance.  Additional ISPI runs for 
$0.18\le \Gamma\delta_t\le 0.22$ and $0.3\le \Gamma\delta_t\le 0.4$ 
were consistent with these results, and
we conclude that small error bars indicate that convergence has been reached.
When dealing with $3^{2K}$ summands in the evaluation of the generating function in the 
AH case as compared to $2^{2K}$ summands for the Anderson model, our ISPI code
to calculate $I(\delta_t,K)$ runs for $\approx 11$~CPU hours on 
a 2.93 GHz Xeon processor. 
\begin{figure}
\centering
\includegraphics[width=\columnwidth]{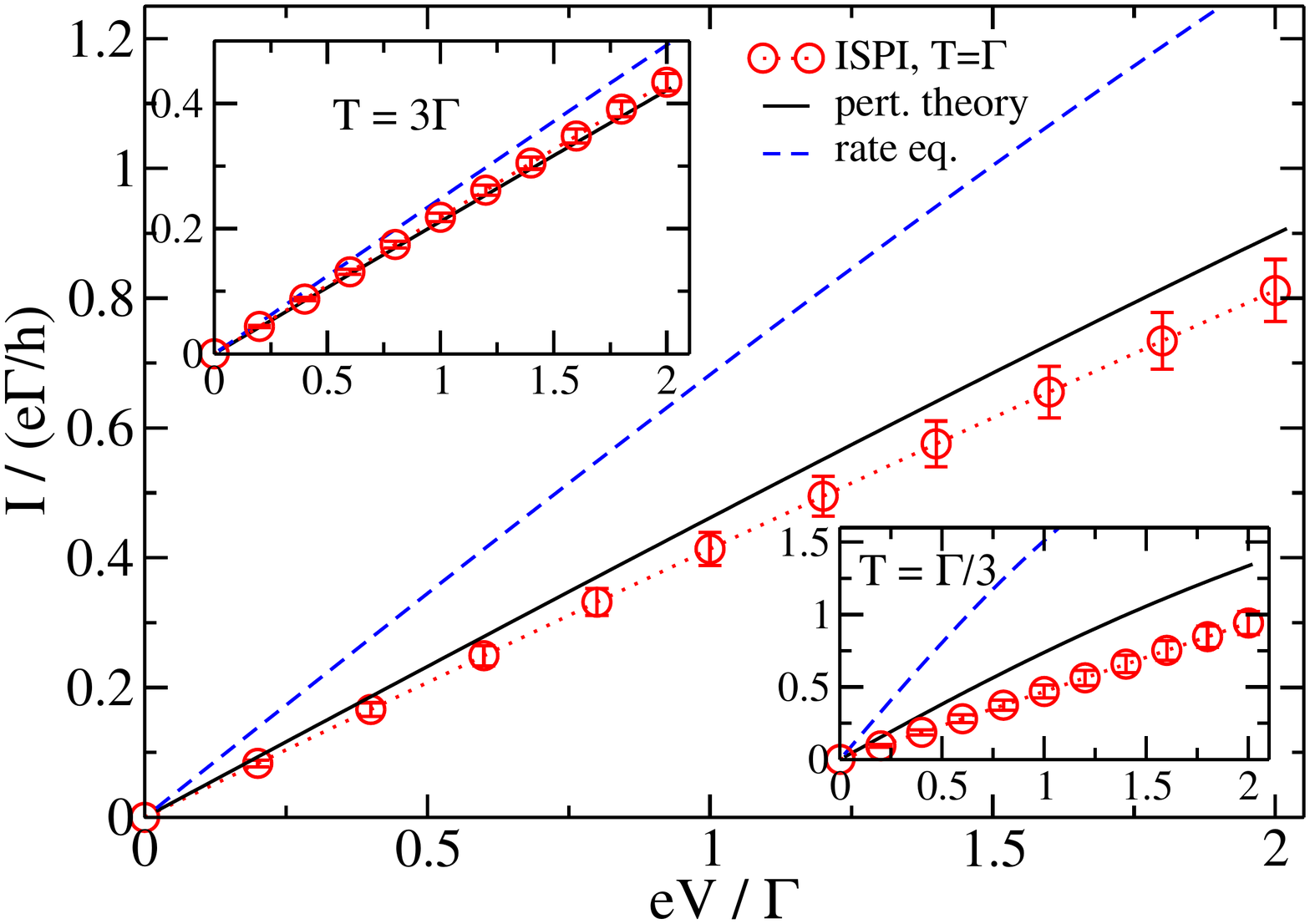}
\caption{(Color online) \label{fig1} 
Current $I$ (in units of $e\Gamma/h$)
vs bias voltage $V$ for the spinless Anderson Holstein model
for 
$\lambda=0.5\Gamma$, $\Omega=\Gamma$, $E_0=0$, and $T=\Gamma$. The
ISPI data are depicted as red circles, where the dotted red curve 
is a guide to the eyes only and the error bars are explained in the main text.
We also show the results of a perturbation theory in
$\lambda$ (solid black curve) and of the rate equation 
(dashed blue curve). The upper (lower) inset shows the 
corresponding result for $T=3\Gamma$ ($T=\Gamma/3$).}
\end{figure}
\begin{figure}
\centering
\includegraphics[width=\columnwidth]{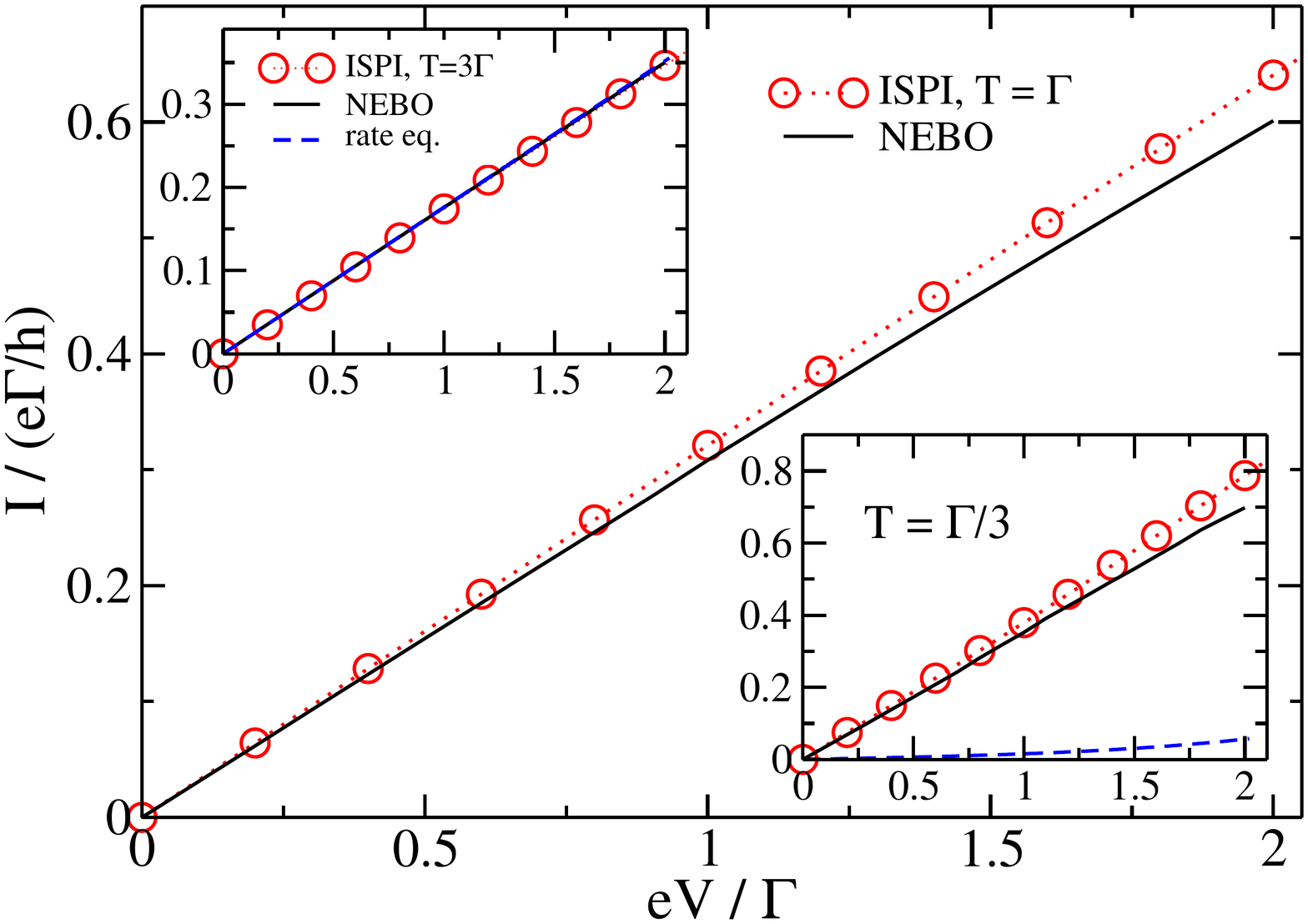}
\caption{(Color online) \label{fig3}
Same as Fig.~\ref{fig1} but for $\Omega=0.5\Gamma$ and
$\lambda=\Gamma$. The main panel is for $T=\Gamma$ and compares the ISPI 
results to NEBO predictions. The insets are for $T=3\Gamma$ and $T=\Gamma/3$, 
respectively, where also the rate equation results are shown. Notice that 
in contrast to ISPI, the rate equation predicts an 
unphysical current blockade for $T=\Gamma/3$. }
\end{figure}

Here, we have once more convinced ourselves that the numerical ISPI results
for the $I(V)$ curves are consistent with known analytical theory in the
respective parameter limits. We employ the following perturbative methods: 
(i)  For $\lambda/\Gamma\ll 1$,
perturbation theory in the electron-phonon coupling applies and 
yields a closed $I(V)$ expression for arbitrary values of 
all other parameters \cite{egger}.  We note that the solution
of the AH model with a very broad dot level \cite{dora,vinkler}
corresponds to this small-$\lambda$ regime.
(ii) For high temperatures, $T\gg \Gamma$, 
a description in terms of a rate equation is possible \cite{nazarov}.  
We here use the sequential tunneling approximation with 
golden rule rates \cite{flensberg}.  For small $\lambda$, the 
corresponding results match those of perturbation theory, while in 
the opposite strong-coupling limit, the Franck-Condon blockade 
occurs and implies a drastic current suppression at 
low bias voltage \cite{ensslin,koch}. 
(iii) For small oscillator frequency, $\Omega\ll {\rm min}(\Gamma,eV)$,
the nonequilibrium Born-Oppenheimer (NEBO) approximation is appropriate and
allows us to obtain $I(V)$ from a Langevin equation for the oscillator 
\cite{pistolesi,bode}.  For small $\lambda$, this approach is also
consistent with perturbative theory, while for high $T$, 
NEBO and rate equation results are found to agree.   For clarity, we
focus on a resonant level with $E_{0}=0$ here.
 The case of weak electron-phonon coupling,
$\lambda=0.5\Gamma$ is shown Fig. \ref{fig1}. We compare our ISPI data for 
$\Omega=\Gamma$ to the results of perturbation theory 
in $\lambda$ and of the rate equation. Perturbation theory essentially
reproduces the ISPI data.
The rate equation is quite accurate for high temperatures, 
but quantitative agreement with ISPI was obtained only 
for $T\gtrsim 10\Gamma$. We note that the ISPI error bars increase when 
lowering $T$ due to the growing memory time ($\tau_c$) demands. 

Next, Fig.~\ref{fig3} shows ISPI results for a slow phonon mode, 
$\Omega=\Gamma/2$, with larger electron-phonon coupling $\lambda=\Gamma$. 
In that case, perturbation theory in $\lambda$ is not reliable and likewise, 
the rate equation is only accurate at the highest temperature ($T=3\Gamma$)
studied, cf.~the upper left inset of Fig.~\ref{fig3}.
However, we observe from Fig.~\ref{fig3} that for such a slow phonon mode, 
NEBO provides a good approximation for all temperatures and/or voltages of 
interest.  We conclude that the ISPI technique is capable of accurately 
describing three different analytically tractable parameter regimes.  

In the limit of strong 
electron-phonon coupling $\lambda$, the classical rate equation  
predicts a Franck-Condon blockade of the current for low bias and $T\gg \Gamma$ 
\cite{koch}. Sufficiently large $\lambda$ can be realized experimentally, and 
the Franck-Condon blockade has indeed been observed in suspended carbon
nanotube quantum dots \cite{ensslin}.  For a nonequilibrated phonon 
 with intermediate-to-large $\lambda$, understanding the Franck-Condon  
blockade in the quantum coherent regime of low temperature, 
$T < \Gamma$, is an open theoretical problem. Here, multiple phonon 
excitation and deexcitation effects generate a complicated (unknown)
nonequilibrium phonon distribution function, and the one-step tunneling 
interpretation in terms of Franck-Condon factors between 
shifted oscillator parabolas \cite{koch} is no longer applicable. 
We here study this question using ISPI simulations, which 
automatically take into account quantum coherence effects.
\begin{figure}
\centering
\includegraphics[width=\columnwidth]{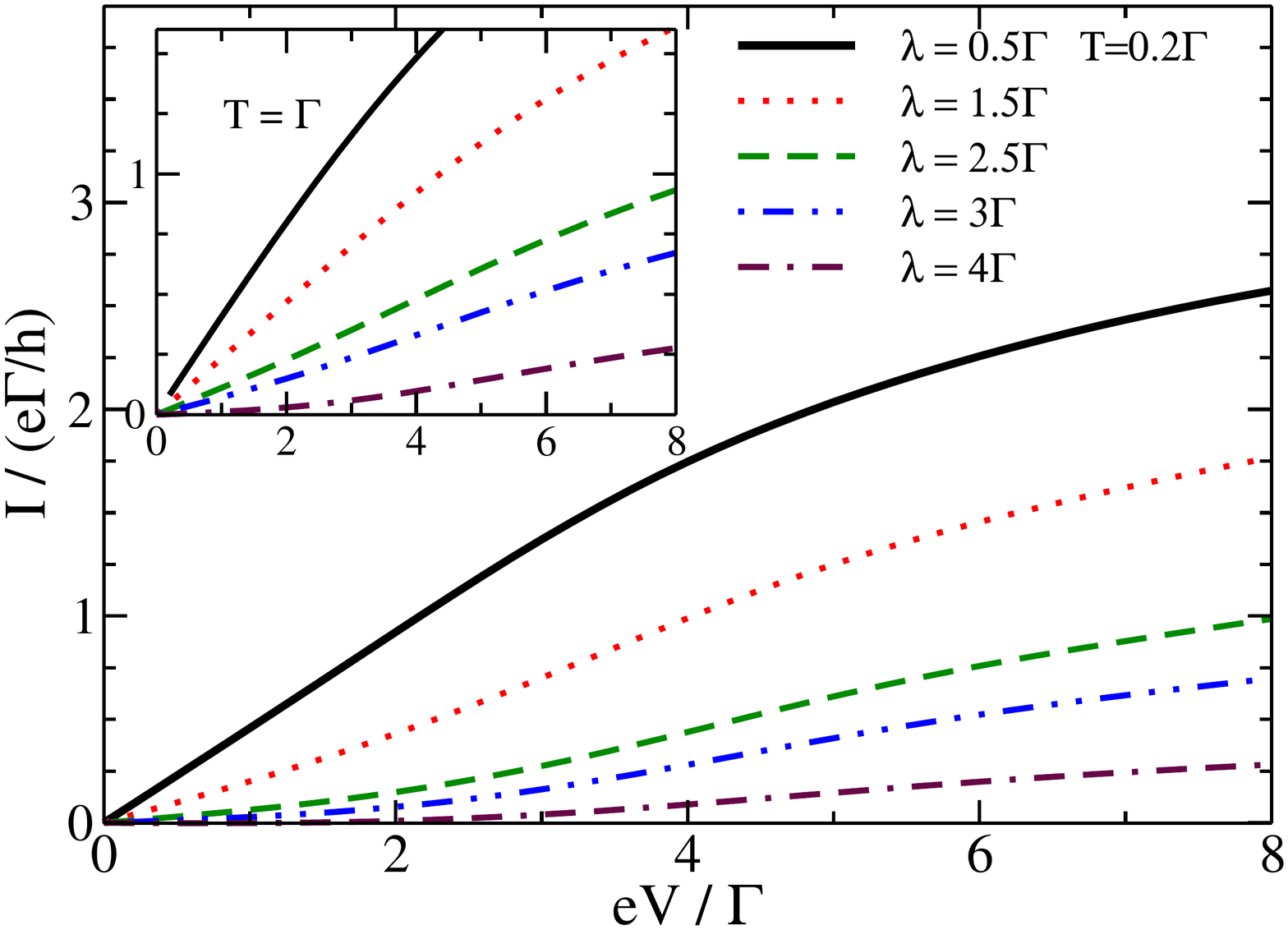}
\caption{(Color online) \label{fig4}
ISPI data for the $I(V)$ curves for the spinless Anderson Holstein model 
from weak ($\lambda=0.5\Gamma$) to strong 
($\lambda=4\Gamma$) electron-phonon coupling, with $\Omega=2\Gamma$. 
The main panel is for $T=0.2\Gamma$, the inset for $T=\Gamma$.
We used a dense voltage grid yielding smooth $I(V)$ curves.
Error bars are not shown but remain small, cp.~Fig.~\ref{fig1}.}
\end{figure}
In Fig.~\ref{fig4}, the crossover from weak to strong electron-phonon coupling 
$\lambda$ is considered. The inset shows $I(V)$ curves for $T=\Gamma$, where we 
observe a current blockade for low voltages once 
$\lambda\gtrsim 2\Gamma$. 
The blockade becomes more pronounced for increasing $\lambda$ 
and is lifted for voltages above the polaron energy
$\lambda^2/\Omega$ \cite{koch}. Remarkably, the Franck-Condon blockade 
persists and becomes even sharper as one enters the quantum-coherent regime 
(here, $T=0.2\Gamma$), despite of the breakdown of the sequential tunneling 
picture. We also observe a nonequilibrium smearing of phonon step-like 
features in the $I(V)$ curves in Fig.~\ref{fig4},
 cf.~also Refs.~\cite{ensslin,koch}.

\section{Nonequilibrium quantum dynamics in the magnetic Anderson model}
\label{magAnderson}
\begin{figure}
\begin{center}
\includegraphics{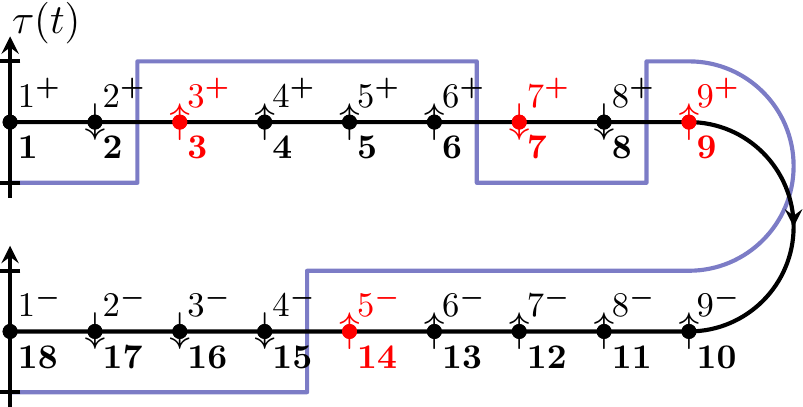}
\caption{
\label{fig:ConstructP}
Exemplary impurity Keldysh path (blue line). The
Keldysh contour is divided into $N - 1 = 8$ segments of length $\delta_{t}$ between $2 N = 18$ time
vertices. The impurity path (tuple of black and red arrows) realizes $m=4$ flip-flops along the contour.}
\end{center}
\end{figure}

The third example to which we have applied the ISPI
scheme \cite{DanielNJP} is the magnetic Anderson model. We extend the Hamiltonian 
Eq.\ \eqref{andham} in the presence of Coulomb interactions, see Eq.~\eqref{HU},  by a magnetic impurity localized on the quantum dot
which interacts via an exchange interaction with the spins of the confined 
electrons on the QD. The magnetic part of the Hamiltonian reads
\begin{eqnarray}
H_{imp} +H^J_{int}=\nonumber\\
\frac{\Delta_{imp}}{2} \tau_z +\underbrace{J \tau_z  (d^\dag_{\uparrow}
d_{\uparrow} - d^\dag_{\downarrow} d_{\downarrow})}_{\displaystyle
H_{\text{int}}^{\parallel}}+ 
\underbrace{\frac{J}{2} (\tau_{+} d^\dag_{\downarrow} d_{\uparrow}  +
\tau_{-} d^\dag_{\uparrow} d_{\downarrow})}_{\displaystyle
H_{\text{int}}^{\perp}}. \nonumber \\
\label{Himp}
\end{eqnarray} 
The  generating function $\mathcal{Z}[\eta]$ is obtained by integrating again 
over the corresponding Grassmann fields for dot and lead operators  
as well as the discrete paths $\{\tau\}$ and $\{\zeta\}$ for the real spin
variables and the HS Ising fields, respectively, and  
\begin{equation}\label{eqn:KeldyshPathIntZ0}
\mathcal{Z}[\eta]=\sum_{\{\tau, \zeta\}}
\int\!\! \mathcal{D} [\bar{c}_{kp\sigma}\bar{d}_\sigma c_{kp\sigma} d_\sigma] (-1)^{\ell} \left(-\frac{i J
\delta_{t}}{2}\right)^{m} P [\{\tau\}] e^{i S }.
\end{equation}
The path sums over impurity and HS 
spin-fields are performed over the $2N$-tuples $\{\tau_j\} = (\tau_{2 N}, \ldots,
\tau_{1})$ and $\{\zeta_j\} = (\zeta_{2 N}, \ldots, \zeta_{1})$ with 
$\tau_j, \zeta_j=\pm 1$.
Within an impurity path $\{\tau\}$, $m$ flip-flop transitions occur on the 
Keldysh contour, where $\ell$ of them lie on the lower branch. The  action $S$
includes tunneling and lead effects  as in Eq.\ \eqref{Seff}. 
Correspondingly, the magnetic part of the action is 
\begin{equation}
S_{\text{imp}}
=-\frac{\Delta_{\text{imp}} \delta_{t}}{2} \sum_{k = 2}^N ( \tau_k  -
\tau_{2 N - k + 1} )
=-\frac{\Delta_{\text{imp}}}{2}
\int_{\mathcal{K}}dt\, \tau (t).
\end{equation}
The polynomial $P[\{\tau\}]$ in Eq.~\eqref{eqn:KeldyshPathIntZ0} depends on the
impurity path $\{\tau\}=\{\tau^+\} (\{\tau^-\})$ for the forward (backward) branch of the contour.
Then, we collect all indices of the flips into the
tuple  $T_{\text{flip}}^{+} = (k^{+}_{m - \ell},\ldots,k^{+}_{1})$ (sorted in
ascending order) along
the forward path $\{\tau^+\} := (\tau_N,\ldots,\tau_1)$ with $\tau_{k^+} \ne
\tau_{k^+ - 1}$ for all $k^+ \in T_{\text{flip}}^+$. Accordingly, 
$T_{\text{flip}}^- = (k^-_{\ell},\ldots,k^-_{1})$ is the tuple of ascending flip
indices along the backward path $\{\tau^-\} := (\tau_{2N},\ldots,\tau_{N+1})$
with $\tau_{k^-} \ne \tau_{k^- + 1}$ for all $k^- \in T_{\text{flip}}^-$. Note
that a flip index on the backward path is labelled according to the
\emph{smaller} step index of the flipping spins corresponding to the
\emph{later} time. The impurity polynomial can be expressed in terms of 
the electronic Grassmann fields as 
\begin{equation}\label{eqn:KeldyshFlipFlopPolynomial}
P [\{ \tau \}]:=\prod_{j \in T_{\text{flip}}^-}  \bar{d}_{\tau_j}^{j +
1} d_{-\tau_j}^{j} \prod_{k \in T_{\text{flip}}^+} 
\bar{d}_{-\tau_k}^{k} d_{\tau_k}^{k - 1}\, .
\end{equation}
Figure \ref{fig:ConstructP} illustrates an example of an impurity path. 
 Collecting all pieces, the remaining formally exact expression for the Keldysh
generating function is
\begin{eqnarray}\label{eqn:KeldyshPathIntZ0poly}
\mathcal{Z}[\eta]&=&\sum_{\{\tau, \zeta\}} \langle P[\{\tau\}]\rangle \prod_{\sigma} \det \{ ( i G_{\sigma}^{\text{eff}} [\{\tau,\zeta\},\eta])^{-1}\}.
\end{eqnarray}
The Keldysh partition function is given as a
sum over expectation values of the polynomial $P$ of Grassmann numbers in a system with Green's function
$G_{\sigma}^{\text{eff}}$. In passing, we note that it is possible to express
the expectation values for the polynomials $P$ in terms of 
Green's functions for the interacting system. The details are given in
Ref.~\cite{DanielNJP}.
After applying Wick's theorem we obtain explicit expressions for the
polynomial \cite{DanielNJP}.
Using matrix elements $(\Xi_{\sigma})_{k, l} =  -i
\langle d_{\sigma}^{q_k} \bar{d}_{\sigma}^{r_l}\rangle  = (G_{ \sigma}^{\text{eff}})_{q_kr_l},$ the final expression for the generating function follows as 
\begin{eqnarray}\label{eqn:FinalGenFunc}
		\mathcal{Z}[\eta]
	&=&	\lim_{\delta_{t} \to 0} \sum_{\{\tau, \zeta\}}
		(-1)^{\ell} \Bigl(\frac{J \delta_{t}}{2 }\Bigr)^{m} \nonumber \\
& & \times \exp \{ i S_{\text{imp}} \}
		\prod_{\sigma}\det i ( G_{\sigma}^{\text{eff}} )^{-1}\det \Xi_{\sigma},
\end{eqnarray}
where the summation over impurity paths is restricted to tuples $\{\tau\}$  with
$\tau_1 = \tau_{2 N} = \tau_i$, i.e., correct boundary conditions along the 
Keldysh contour are fulfilled. The limit $\delta_t\to 0$ appears explicitly
here, since there is no continuous measure used for the discrete spin paths,
neither for the HS- nor for the impurity spins.
In order to reduce the exponentially growing number of contributing paths 
($\sim 4^K$, due to real spin and HS spins)
without affecting the accuracy, we may exploit 
that the propagating tensor depends on the number $m_j$ of flip-flops in  path
segment $j$ and $0 \le m_{j} \le 2
(K - 1)$ along the Keldysh contour. We observe that the
weight of each segment is smaller, the more flip-flops it contains. 
On the other hand, the number of path segments
$\{\tau\}_{j}$ with $m_{j}$ flip-flops (given by $4 C^{2 (K -1)}_{m_j}$ with
$C^n_{k} = n! / [k! (n - k)!]$) grows as long as $0\le m_{j}\le K - 1$, 
{\em but decreases again} when $K \le m_{j} \le 2 (K - 1)$. As a consequence,
for any observable there exists a maximal $m_j^{max}$ such that contributions from paths with 
$m_j>m_j^{max} \le 2(K-1)$ could safely be disregarded in the numerical iteration. 
Of course $m_j^{max}$ is chosen depending on 
the model parameters and the observable under investigation.

Rapidly decreasing weights of the paths
may not be (over-)compensated by their increasing numbers for $0 \le m_{j} \le K
- 1$, since each contribution is small and the number
of paths decreases again for larger $m_{j}\ge K$.
The behavior of the impurity weights is illustrated as follows. Consider 
the case when $m_j$ is close to the maximum $2 (K - 1)$. Both path
classes with $m_{j} = 0$ and $m_{j} = 2 (K - 1)$ contain the same number of
elements (four), while each path contribution in the second class
is weighted by $(J \delta_{t} / 2)^{2 (K - 1)}$. For typical values of $K =
4$, $\delta_{t} \Gamma = 1/2$, and $J = \Gamma$, the weight is $\sim 2.5 \times 10^{-4}$. This
also holds for all $K \le m_{j} \le 2 (K - 1)$. Since $m_{j}^{\text{max}}$ is 
unknown {\em a priori\/}, we include it into our code as an additional parameter.
Then, we perform a numerical estimate 
by a spot sample of the parameter space. It turns out that for the considered 
cases, it is sufficient already to choose $m_{j}^{\text{max}} = 2$. 
This drastically reduces the CPU running times from more than one month to typically three to five days.

\subsection{Impurity dynamics} \label{sec:results}
We focus on transport features caused by the magnetic impurity in this section.
\begin{figure}
\begin{center}
\includegraphics[width=\columnwidth]{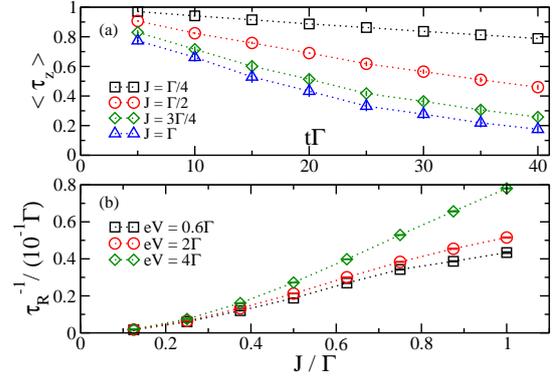}
\end{center}
\vskip-\lastskip
\caption
{\label{fig:TimeDecayOfImpSpin} (a) The expectation value $\langle \tau_z\rangle$ of the impurity orientation as a function of propagation time for different strengths $J$ of an anti-ferromagnetic electron-impurity interaction. The initial preparation of the system at $t=-\infty$ is spin-up [$\tau_z (0) \equiv \tau_i = 1$], for the other parameters see text. The polarization decays faster with increasing $J$. (b) Impurity relaxation rate $\tau_{R}^{-1}$ as a function of the exchange coupling strength. Again the relaxation is faster with increasing coupling strength. In both panels the temperature is $T=\Gamma$.}
\end{figure}
We emphasize that novel dynamical and transport features are mediated by the
transverse or flip-flop interaction $H_{\text{int}}^{\perp}$,  given in
Eq.~\eqref{Himp}. Without the possibility of flip-flops the orientation of the
impurity spin and its quantum state could not change. 
The remaining longitudinal part $H_{\text{int}}^{\parallel}$ of the interaction 
causes a renormalization of rates and energies which appears as effective
magnetic field.
Necessarily, flip-flop processes are
involved from the beginning to investigate the non-trivial impurity dynamics 
by considering the time dependence of the impurity orientation $\langle \tau_z\rangle$. 
In all presented results below,  the impurity is fully polarized at $t=-\infty$
and the coupling to the leads is switched on. 

In Fig.~\ref{fig:TimeDecayOfImpSpin} (a) we present the time evolution of the impurity polarization
 $\langle \tau_z\rangle $ for different values of the exchange interaction $J$. 
The remaining parameters are $\Phi_D = \Delta =
\Delta_{\text{imp}} = U = 0$, and $T=\Gamma$ and $e V = 0.6
\Gamma$. The impurity polarization shows a clear exponential decay $
\langle \tau_z\rangle  (t) \propto e^{-(t - t_i)\tau_{\text{R}}^{-1}}
$,
well described by a single relaxation rate for intermediate to long
propagation times.
A faster decay is observed as the impurity interacts stronger
with the electron spins. The parameters are chosen to yield an isotropic
(symmetric with respect to [relative] spin orientations) model system. In this
case the antiferromagnetic interaction favors antiparallel orientation of
electron- and impurity spin. Over long propagation times, the coupling to the
unpolarized leads then destroys any polarization of the impurity. It is
therefore reasonable to assume, that the rates for up- and down flips are
equal. 
While the impurity interaction energy is comparable to the tunneling coupling
and considerably affects the transport behavior as we show below (see
Fig.~\ref{fig:USweepCurrent}), the rather high temperature and bias voltage
nevertheless reduce the relevance of coherent dynamics due to on-dot
interactions to a secondary role.

 We next investigate the relaxation rate $\tau_{\text{R}}^{-1}$. In Fig.~\ref{fig:TimeDecayOfImpSpin}(b) for $T=\Gamma$. We present results for varying $J$ and $U
= 0$, and three different bias voltages. These show a nearly quadratic behavior 
growing from zero (no relaxation) in the sense that for a fit of the results for
$0 \le J \le \Gamma/2$ to a polynomial function $a J^b$ the exponent $b$ lies
between $\sim 1.8$ and $\sim 1.9$.
An \emph{exact} quadratic dependence of $\tau_{\text{R}}^{-1}$ on $J$ is 
obtained only when the dynamics is \emph{strongly} dominated by
sequential (incoherent) flip-flop processes \cite{DanielNJP}. This is only realized when $J \ll
\Gamma$. A sequential flip-flop process consists of three elementary 
components: the actual flip-flop and two tunnelling processes of single
electrons with opposite spins (not necessarily in that order). Since they evolve
coherently, these components form an effective spin-flip process $|\chi,
\tau\rangle \to |\chi, -\tau\rangle$, where $\chi \in \{0, \sigma, \text{d}\}$
and the underlying flip-flop nature is masked by the tunnelling electrons.  

\subsection{Charge current for finite impurity interaction and Coulomb repulsion}
In the deep quantum regime, where no small parameter exists, ISPI is able to describe physical properties not predictable by
perturbative methods. In this section, we study how the current behaves as
functions of bias voltage, Coulomb interaction and temperature, respectively. 
\begin{figure}
\begin{center}
\includegraphics[width=\columnwidth]{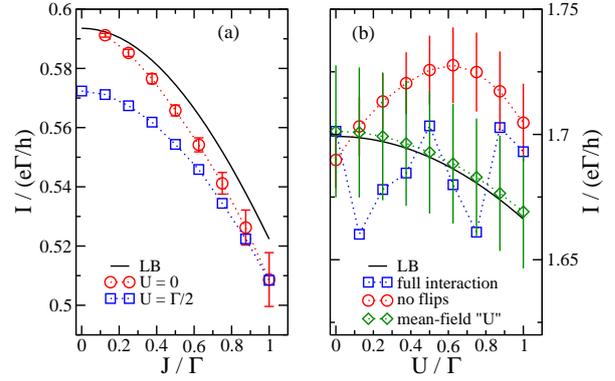}
\end{center}
\caption
{\label{fig:USweepCurrent}
(a) Charge current as a function of the exchange interaction $J$ for two values
of the on-dot interaction $U$ at $T=\Gamma$ and $eV=0.6\Gamma$. The solid line marks the Landauer-B\"uttiker result, where exchange correlations are treated on a mean field level, see Ref.~\cite{DanielNJP}. 
(b) Comparison of (i) the LB current (solid lines), (ii) the Coulomb interacting current without flip-flop scattering (``no flips'', red circles), (iii) the current without Coulomb scattering but full impurity interaction (``mean-field $U$'', green diamonds, see the text for explanation), and (iv) the fully interacting current (``full int.'', blue squares) in their dependence on the Coulomb interaction $U$ for $T=\Gamma$ . The other (non-zero) parameters are $J = \Gamma$ and $eV = 2 \Gamma$.}
\end{figure}

Fig.~\ref{fig:USweepCurrent} (a) shows that the flip-flop term
$H_{\text{int}}^{\perp}$ has a considerably smaller influence on the charge
current at $T=\Gamma$ (incoherent regime) than the
longitudinal part of the interaction in Eq.~\eqref{Himp}. Despite the qualitatively similar behavior
of the Landauer-B\"uttiker (LB) current and the exact data, the flip-flop scattering causes an
additional significant current drop that grows for growing $J$. A finite Coulomb
interaction of $U = \Gamma / 2$ increases the resistivity of the dot and the ISPI
points are consistently lower than the LB values. The voltage is chosen as $eV=0.6\Gamma$.
In Fig.~\ref{fig:USweepCurrent} (b), for $T=\Gamma$  four different current
curves are shown---one for each possibility to either have (i) only mean field
dynamics, regarding $J$ (LB), (ii) the full Coulomb interaction without flip-flop
processes (``no flips''), (iii) flip-flop dynamics without Coulomb fluctuations
(``mean-field $U$''), and (iv) the fully interacting dot (``full int.''). For $J =
\Gamma$ and $V = 2 \Gamma$, the Coulomb energy is varied between $0 \le U \le
\Gamma$. The situation ``mean-field $U$'' is implemented by setting $\Phi_D=U / 2$ and
the HS parameter $\lambda=0$ to illustrate the effect of the ``classical'' part
of the Coulomb interaction. Only for the ``single-interaction'' currents (``no
flips''), we show the error bars. We do not show a margin of confidence for the
fully interacting case in order that the error
data remain comparable. Calculating the ``full int.'' current is a time
consuming task and thus, the extrapolation involves considerably fewer data
points. Nevertheless, this does not render these values unreliable (we still see
a compelling linear behavior of the $1/\tau_{c}$ extrapolation with errors of
the order of $1\%$ based on the sample standard deviation). Both the mean-field current and the current without Coulomb
scattering show only a weak dependence on $U$ due to the single-particle energy
shift. The current with full Coulomb interaction but fixed impurity shows a
local maximum for $U \sim \Gamma / 2$. In this case, the fixed impurity acts as
an effective static magnetic field. The ISPI values for the fully interacting
dot vary strongly over the considered $U$ interval, but are scattered around the
``no flips'' and ``mean-field $U$'' curves.

As long as the Coulomb interaction is small, differences in the data arise 
from including or excluding flip-flop processes. Hence, the rather good
agreement of the $U = 0$ values suggests that even at this, temperature
flip-flop processes alone affect the current only weakly. Nevertheless, for
decreasing temperatures, the flip-flop processes start to influence the current
more strongly, which results in an increased resistivity. The case of the ``no
flip'' current (fixed impurity) is equivalent to a Coulomb-interacting
single-level quantum dot in a magnetic field. This effect is caused by the
broadening of the dot's joint density of states due to the Coulomb fluctuations.

\section{Conclusions}
\label{conclusion}

In summary, we have reviewed a scheme for the iterative summation 
of real-time path integrals (ISPI) and applied it to
 prototypical problems of quantum transport 
through an interacting quantum dot coupled to 
metallic leads held at different chemical  
potentials. After integrating over the 
leads' degrees of freedom, a time-nonlocal Keldysh self-energy 
arises. Exploiting the 
exponential decay of the time correlations at finite temperature 
allows us to introduce a memory time $\tau_c$ 
beyond which the correlations can be truncated. 
Within $\tau_c$,  correlations are fully taken into account 
in the corresponding path integral for the Keldysh generating function. Then, 
through a discrete Hubbard-Stratonovich transformation, interactions are
transferred to an auxiliary spin field, and 
an iterative summation scheme is constructed. The remaining systematic
errors due to the finite time discretization and the finite memory time
$\tau_c$ are eliminated by a refined Hirsch-Fye-type 
extrapolation scheme, rendering the ISPI numerically exact. 

The scheme has been applied to the canonical
 example of a single-impurity Anderson dot with Coulomb 
interaction $U$. This allows us to carefully and systematically check
the algorithm. For linear transport, we have recovered results 
from second-order perturbation theory in $U$ in the limit of very
small interaction strength, but found significant deviations already for
small-to-intermediate values of $U$. In the incoherent sequential regime, 
we recover results from a master equation approach. We have furthermore
reproduced the linear conductance above the Kondo temperature.
 In addition, we have investigated the regime of 
correlated nonlinear transport, where, in our opinion, the 
presented method is most valuable.  
The {\em nonequilibrium\/} Kondo regime, representing 
an intermediate-to-weak coupling
 situation, seems tractable  by the ISPI scheme.

Our approach is, in fact, similar in spirit to the
well-established concept of the
quasi-adiabatic path integral (QUAPI) scheme, introduced by Makri and Makarov 
\cite{Makri} in its iterative version. This method has been  developed to describe the dynamics 
of a quantum system coupled to bosonic environments, see also Refs.~\cite{Eckel,Thorwart1}.

In a second example, we have applied the ISPI technique to the spinless Anderson-Holstein
model as well, which is 
the simplest nonequilibrium model for  molecular quantum dots  
with a phonon mode. Our formulation exploits a mapping 
to an effective three-state system and reproduces three analytical 
theories valid in different parameter regions. This extension of the 
ISPI approach then captures the full crossover between those limits.  For
strong 
electron-phonon coupling and a nonequilibrated phonon mode, we find that
the Franck-Condon blockade becomes even more pronounced 
as one enters the deep quantum coherent regime.  

The complex system of an incorporated spin-1/2 magnetic impurity in a quantum
dot has been subject of a third study, focusing
on the real-time dynamics in the presence of Coulomb interactions. We 
include the impurity interaction on the same level as	 the other interactions,
which results in an additional sum over impurity paths. 
An efficient truncation scheme nevertheless provides
accurate results for the coupled spin dynamics. Results are given for a quantum
spin-$1/2$ impurity on the dot, whereas the generalization to an impurity with a
larger spin is possible. For a small impurity interaction, where sequential
flip-flops dominate the impurity
dynamics, we have found good agreement with a classical rate equation, see Ref.~\cite{DanielNJP} as well. This is a
useful tool to gain insight into the dominating processes in the incoherent
regime. Relaxation is described reasonably well by a rate equation when
lead-induced coherences are absent. 

In the deep quantum regime, however, we find that the ISPI method is the only
tool to obtain both the correct order of magnitude and the qualitative features
of the relaxation rate as it depends on the system parameters $U$ and $J$. The
same holds for the influence of $J$ and $U$ on the current in this interesting corner of the
parameter space.  Most importantly, the ISPI scheme proves to be useful to cover the full
cross-over regime where no small parameter exists and thus any
perturbative approach becomes invalid. 

We have provided a first glimpse on the interesting new physics that comes into
reach with the ISPI scheme. Compared to other approaches, it has several 
advantages (e.g., numerical
exactness, direct nonequilibrium formulation, no sign problem), 
but is, on the other hand,  computationally more costly than most
other techniques, especially for strong correlations and/or low energy
scales (temperature, voltage).

\begin{acknowledgement}
Financial support by the DFG through SPP 1243 ``Quantum transport  at the
molecular scale'',  DFG SFB 668 ''Magnetismus vom Einzelatom zur
Nanostruktur'' and DFG project KO $1987/5$ (SW) is acknowledged.  Computational time from the ZIM at
Heinrich-Heine Universit\"at D\"usseldorf with support from S. Raub is greatly acknowledged. 
\end{acknowledgement}

\end{document}